\documentclass[12pt]{article}
 \usepackage{amsmath,amsxtra,amssymb, amsfonts }

  \numberwithin{equation}{section}

\newtheorem{thm}{Theorem}
\newtheorem{cor}[thm]{Corollary}
\newtheorem{lemma}[thm]{Lemma}

\def\be{\begin{eqnarray}}
\def\ee{\end{eqnarray}}
\def\bee{\begin{eqnarray*}}
\def\eee{\end{eqnarray*}}
\def\bal{\begin{align}}
\def\enal{\end{align}}
\def\pmx{\begin{pmatrix}}
\def\emx{\end{pmatrix}}
\def\bsq{\begin{subequations}}
\def\esq{\end{subequations}}

\def\ds{\displaystyle}
\def\ts{\textstyle}
\def\nn{\nonumber}
\def\half{{ \frac{1}{2} }}

\def\raw{\rightarrow}
\def\nrm{\|}
\def\ot{\otimes}

\def\eps{\epsilon}

\def\tr{{\rm Tr} \, }
\def\trp{{\rm Tr} }
\def\bra{\langle}
\def\ket{\rangle}
\def\kb{ \ket \bra }
\def\wtd{\widetilde}
\newcommand{\proj}[1]{ | #1 \kb  #1|}
\newcommand{\norm}[1]{ \| #1  \|}
\newcommand{\bnorm}[1]{ \big\| #1  \big\|}

\def\mm{ \! - \!}
\def\pp{ \! + \!}

\def\haf{{ \tfrac{1}{2} }}
\def\haf{ \frac{1}{2} }
\def\rt2{\ts \frac{1}{\sqrt{2}} }
\def\ovb{\overline}

\def\hil{{\cal H}}
\def\id{{\cal I}}
\def\rmi{{\rm I}}
\def\qed{\qquad{\bf QED}}
\def\cbm{{\rm CB,min}}
\def\cbqp{{\rm CB,} q \raw p}
\def\cb1p{{\rm CB,} 1 \raw p}

\def\prf{\noindent {\bf Proof: }}

 \newcommand{\pf}[1]{\noindent {\bf Proof of   #1: }}
\newcommand{\rmk}{\noindent {\bf Remark: }}

\def\cd{{\bf C}^d}

\title{Multiplicativity of completely bounded p-norms
implies a new additivity result}

   \author{Igor Devetak \\
    {\small   Electrical Engineering Department,  University of Southern California, Los Angeles CA 90089} \\
     {\small devetak@csi.usc.edu}\\ \and Marius Junge \\  
{\small Department of Mathematics, 
    University of Illinois at Urbana-Champaign, Urbana IL 61801} \\  
     {\small mjunge@math.uiuc.edu} \\
    \and Christopher King \\
{\small Department of Mathematics,
Northeastern University,  Boston MA 02115} \\
 {\small  king@neu.edu} \\
\and Mary Beth Ruskai
      \\  {\small  Department of Mathematics,
   Tufts University,
     Medford MA 02155} \\ 
   {\small Marybeth.Ruskai@tufts.edu}}
     
\begin{document}
     
     \maketitle
     
     \begin{abstract}
   We prove additivity of the minimal conditional entropy associated
   with a quantum channel $\Phi$, represent by a completely positive
   (CP), trace-preserving map, when the infimum of
   $S(\gamma_{12}) - S(\gamma_1)$ is restricted to states
   of the form $(\id \ot \Phi)\big( | \psi \kb \psi | \big)$
   We show that this follows from multiplicativity of the
   completely bounded norm of $\Phi$ considered as
   a map from $L_1 \raw L_p$ for $L_p$ spaces defined
   by the Schatten p-norm on matrices, and give another proof
   based on entropy inequalities.  Several related
   multiplicativity results are discussed and proved.
   In particular, we show that both the usual $L_1 \raw L_p$ 
   norm of a CP map and the  corresponding completely bounded
   norm are achieved for positive semi-definite matrices.
   Physical interpretations are considered, and a new proof 
   of strong subadditivity is presented.
     \end{abstract}
\tableofcontents
 \section{Introduction}  \label{sect:intro}

Quantum channels are represented by 
completely positive, trace preserving (CPT) maps
on $M_d$, the space of $d \times d$ matrices.
Results and conjectures about additivity and superadditivity of
various types of capacity play an important role in
quantum information theory.   

In this paper, we present a new additivity result which
can be stated in terms of a type of minimal conditional
entropy defined as
\be  \label{cbmdef}
   S_{\cbm}(\Phi) = \inf_{\psi \in \cd \ot \cd} \Big(
        S\big[(\id \ot \Phi)\big( | \psi \kb \psi | \big)] - 
          S\big[(\trp_2 \big( | \psi \kb \psi | \big)] \Big)
\ee
where $S(Q) = - \tr Q \log Q$ is the von Neumann entropy.
The shorthand CB stands for ``completely bounded'' which
will be explained later.
We will show that this CB  minimal conditional
entropy is additive, i.e.,
\be   \label{cbmadd}
   S_{\cbm}(\Phi_A \ot  \Phi_B) = S_{\cbm}(\Phi_A) + S_{\cbm}(\Phi_B) .
\ee

The expression \eqref{cbmdef} for $S_{\cbm}(\Phi) $ should be compared to those for two
important  types of capacity.    To facilitate this, it is useful to let
$   \gamma_{12} = (\id \ot \Phi) \big(|\psi \kb \psi| \big)$, and observe that its
reduced density matrices are 
$   \gamma_1 = \trp_2 (\id \ot \Phi) \big(|\psi \kb \psi| \big) $, and 
$   \gamma_2 = \trp_1 (\id \ot \Phi) \big(|\psi \kb \psi| \big)$.
We can now  rewrite \eqref{cbmdef} as
\begin{align}  \label{cbm2}
  - S_{\cbm}(\Phi) & \, = \,      \sup_{\psi} \Big[ S(\gamma_1)  - S(\gamma_{12}) \Big] .
\\ 
    \intertext{The capacity of a quantum channel for transmission
 of classical information when assisted by unlimited entanglement (as in, 
 e.g., dense coding) is given by \cite{BSST1,BSST2,Hv6}}
  C_{EA}(\Phi) & \, = \,     \sup_{\psi} \Big[ S(\gamma_1) + S(\gamma_2) - S(\gamma_{12}) \Big] .
 \label{eacap}  \\ 
    \intertext{The capacity for transmission of
quantum information without additional resources is the  
coherent information, 
 \cite{BNS,Dev,Lloyd,Shor4}}
   \label{Qcap} 
    C_Q(\Phi) & \, = \,       \sup_{\psi} \Big[  S(\gamma_2) - S(\gamma_{12}) \Big]  
     \end{align}
In these expressions, the  supremum is taken over
all  normalized vectors 
$ \psi$ in  $\cd \ot \cd$ and  $\gamma_{12}$  depends on both $\psi$ and $\Phi$.
It has been established that $C_{EA}(\Phi)$ is additive \cite{BSST1,Hv6}, 
but $C_Q(\Phi)$ is not additive in general \cite{DSS}.   To understand the
 difference between $C_Q(\Phi)$ and
$S_{\cbm}(\Phi) $,  use the trace-preserving property of $\Phi$ to rewrite
$\gamma_1= \trp_2\big(|\psi \kb \psi| \big)$ and 
$\gamma_2 =    \Phi \big[ \trp_1\big(|\psi \kb \psi| \big) \big]$. 
The additive quantity  \eqref{cbm2}  contains  $\gamma_1$
 which is independent of  $\Phi$, while the non-additive quantity
  \eqref{Qcap} contains $\gamma_2$  which depends upon $\Phi$.
 
We do not have a completely satisfactory physical interpretation of the the CB entropy,
although an operational meaning can be found.
It appears to provide a measure of how well a channel preserves
entanglement.   In particular, if  $\Phi$ is entanglement breaking, $S_{\cbm}(\Phi) > 0$
(although the converse does not hold).
Recently, Horodecki,  Oppenheim and  Winter \cite{HOW} gave an
elegant interpretation of quantum conditional information which we discuss in
the context of our results in Section~\ref{sect:appl}.

The additivity \eqref{cbmadd}  will follow from the multiplicativity 
 \eqref{mult.omega} of the quantity
\be  \label{cb1p}
   \omega_p(\Phi) \equiv \sup_{\psi \in \cd \ot \cd} 
      \frac{ \nrm (\id \ot \Phi)\big( | \psi \kb \psi | \big) \nrm_p}
      { \nrm \trp_2  \big( | \psi \kb \psi | \big) |_p} = 
      \sup_{\psi \in \cd \ot \cd}  \frac{ \norm{ \gamma_{12}}_p}{ \norm{ \gamma_1}_p} ~.
\ee
We will see that this  is a type of CB  norm.
Recall that one of several equivalent criteria for a map $\Phi$
to be completely positive is that  for all integers $d$, the map $\id_d \ot \Phi$ takes 
positive semi-definite matrices to positive semi-definite matrices.
(We use $\id $ to denote the identity map  $\id(\rho) = \rho$ to avoid
confusion with the  identity matrix $\rmi$.)
One can similarly define other concepts, such as completely isometric,
in terms of the maps $\id_d \ot \Phi$.    The completely bounded (CB) norm
is thus
\be   \label{cbnorm}
   \norm{ \Phi}_{\rm CB} = \sup_{d} \norm{ \id_d \ot \Phi}.
\ee
However, this depends on the precise definition of the norm on the right side
of \eqref{cbnorm} or, equivalently, on the norms used to regard $\Phi$ and
 $\id_d \ot \Phi$ as maps between Banach  spaces.   The appropriate 
 definitions for the situations
considered here are described in Sections~\ref{sect:CBdefs}
and \ref{sect:op.space}.

In the process of deriving our results, we obtain a number of related results
of independent interest.    For example, we show that when $\Phi$ is a CP
map, both  $\norm{\Phi}_{q \raw p}$ and the corresponding CB norm are
attained for a positive semi-definite matrix, extending a result in \cite{Wat}.
  The strong subadditivity (SSA)
 inequality \cite{SSA,RuskJMP} for quantum entropy 
  \be  \label{SSA}
     S(Q_{123}) + S(Q_3) \leq   S(Q_{ 23}) + S(Q_{13}) 
  \ee
  is the basis for Holevo's proof of additivity of $C_{EA}(\Phi) $ and the  
proof of  \eqref{cbmadd} given in Section~\ref{sect:igor}.   In Section~\ref{sect:ent}
we use operator space methods  to obtain a new proof of SSA.

This paper is organized as follows.
Section~\ref{sect:addcb} is concerned with our main result, \eqref{cbmadd}.
After some background, we present two different  proofs.
In Section~\ref{sect:CB}, which is divided
into six subsections,  we introduce notation and summarize  results about 
CB norms and operator
spaces  used in the paper.      Only the basic notation in
Section~\ref{sect:CBdefs} and the
Minkowski inequalities in Section~\ref{sect:fubmink} are needed for the main
result, Theorem~\ref{thm:multqp}.
  A subtle distinction between the norms used to define
$\norm{ \Phi}_{\rm CB}$  and $\norm{ \id_d \ot \Phi}_{q \raw p}$ often used
in quantum information (e.g., \cite{AHW,KR3,Kv,Wat}) is described in the 
penultimate paragraph of Section~\ref{sect:3.2}.
 
 In Section~\ref{sect:CBmult}, we prove multiplicativity of the CB norm 
 for maps $\Phi : L_q(M_m) \mapsto L_p(M_n) $.  
   When  $q \geq p$, we also show that the CB norm 
 equals $\norm{\Phi}_{q \raw p}$, yielding a  proof of  multiplicativity
 for the latter.    In Section~\ref{sect:appl}, we explicitly give $\norm{ \Phi}_{\rm CB}$
 and $S_{\cbm}(\Phi) $ for simple examples, including the depolarizing
 channel; prove that $S_{\cbm}(\Phi) > 0$  for EBT maps; and discuss
 physical interpretations.    In Section~\ref{sect:ent},   we use  
  the Minkowski inequalities for the CB norms to obtain 
  a new proof of SSA.   We also show that the minimizer implicit in
  $\norm{X_{12}}_{(1,p)}$ converges to $X_1$.

 \section{Additivity of CB entropy}  \label{sect:addcb}
 
\subsection{Multiplicativity questions in quantum information theory}  \label{sect:multQ}

We are interested in  CB norms
when $\Phi$ is a map $L_q(M_d) \mapsto L_p(M_d)$ where  
$L_p(M_d)$ denotes the Banach space of $d \times d$
matrices with the Schatten norm 
$ \norm{A}_p =  \big( \tr |A|^p \big)^{1/p}$.   One then defines
the norm
\be   \label{qpnorm}
   \norm{\Phi}_{q \raw p} \equiv  \sup_{A} \frac{ \norm{ \Phi(A)}_p}{\norm{A}_q}
\ee
Watrous \cite{Wat}  and Audenaert \cite{Aud} independently showed that this norm is 
unchanged if the supremum in \eqref{qpnorm} is restricted to positive semi-definite
matrices, resolving a question raised in \cite{KR3}.   Thus,
\be   \label{qpnorm+}
   \norm{\Phi}_{q \raw p} =  \sup_{A > 0} \frac{ \norm{ \Phi(A)}_p}{\norm{A}_q}
\ee

In quantum information theory, the norm  $\nu_p(\Phi) = \norm{\Phi}_{1 \raw p}$ 
  plays an important
 role.   It has been conjectured \cite{AHW} (see also \cite{KR3})  that 
 \be  \label{mult.nup}
 \nu_p(\Phi_A \ot \Phi_B) = \nu_p(\Phi_A) \, \nu_p(\Phi_B) 
 \ee
 in the range $1 \leq p \leq 2$.   Proof
 of this conjecture would imply additivity of minimal entropy
 which has been shown to be equivalent to several other
 important and long-standing conjectures \cite{Shor}.
 We note here only that $\ds{S_{\min}(\Phi) = \inf_{ \rho \in {\cal D}} S[\Phi(\rho)]}$
 where  ${\cal D} = \linebreak  \{ \rho : \rho > 0, ~ \tr \rho = 1\}$
 denotes the set of density matrices.  Note that
 $\ds{\nu_p(\Phi) = \sup_{\rho \in {\cal D}} \norm{ \Phi(\rho)}_p}$.
 Amosov, Holevo and Werner \cite{AHW} showed that the additivity of minimal entropy 
 \be
    S_{\min}(\Phi_A \ot \Phi_B) =   S_{\min}(\Phi_A ) +   S_{\min}( \Phi_B) 
 \ee
would follow  if  \eqref{mult.nup} can be proved.
 
  In this paper, we  consider instead  $\norm{\Phi}_{CB, 1 \raw p}$
 for which the expression in \eqref{cbnorm} reduces to $\omega_p(\Phi)$,
 and show that it is multiplicative, i.e., that
 \be  \label{mult.omega}
    \omega_p(\Phi_A \ot \Phi_B) = \omega_p(\Phi_A) \, \omega_p(\Phi_B) .
 \ee
 We first show that \eqref{mult.omega} implies our new additivity result,
 providing a motivation for the technical material needed to prove 
 \eqref{mult.omega}.   We subsequently found another proof
 which does not use CB norms; this is presented in Section~\ref{sect:igor}.
However, the CB proof given next  provides an
indication of the potential of  this machinery for quantum information.

 \subsection{Proof of additivity from CB multiplicativity}  \label{sect:CBpf}

We define a function of a self adjoint matrix  with
spectral decomposition $A = \sum_k \lambda_k  \proj{\phi_k} $
as $f(A) = \sum_k   f(\lambda_k)  \proj{\phi_k} $.
We will need functions of the form $f(t) = t^p \log t $ defined
on $[0,\infty)$ so that $f(0) = 0$ for $p > 0$ and 
 $Q^p \log Q$ is $0$ on $\ker(Q)$.  
For any $Q > 0$ we define the entropy as $S(Q) = - \tr Q \log Q$
and note that $S\big( \frac{Q}{\tr Q} \big) = \frac{1}{ \tr Q} S(Q) + \log \tr Q $.

We will often use the notation $\gamma_{12}$ for  density matrices in  the tensor
product $M_d \ot M_n \simeq M_{dn}$ and $\gamma_1 = \trp_2 \, \gamma_{12}$,  
 for the corresponding reduced density matrix in $M_d$.
 (The partial trace   $\trp_2$ denotes the trace on   $M_n$.     One
 can similarly define $\gamma_2 = \trp_1 \, \gamma_{12} $
 The density matrix $\gamma_{12}$ can be regarded
 as a probability distribution on ${\bf C}_d \ot  {\bf C}_n$
 in which  case $\gamma_1$ and $\gamma_2$ are  the non-commutative analogues
of its marginals.)  We  first prove a technical result.
\begin{lemma}  \label{lemm:df1}
The function $u(p,\gamma_{12}) \equiv \frac{1}{p- 1} \Big( 1 - 
     \frac{ {\rm Tr}_{12} \, \gamma_{12}^p}{ {\rm Tr}_1 \, \gamma_1^p}  \Big) $ is well-defined
     for $p > 1$ and $\gamma_{12} $ a density matrix.  It can be extended
     by continuity to $p \geq 1$ and this extension satisfies
     \be  \label{df1}
    u(1,\gamma_{12}) = - \frac{d~}{dp} \,  
       \frac{ {\rm Tr}_{12} \, \gamma_{12}^p}{ {\rm Tr}_1 \, \gamma_1^p} \Big|_{p = 1}
      = S(\gamma_{12}) - S(\gamma_1) .
     \ee 
  Moreover,  $u(p,\gamma_{12}) $ is uniformly bounded in $\gamma_{12}$ for
  $p \in [1,2]$ and the continuity at $p = 1$ is uniform in $\gamma_{12}$.
  \end{lemma}
\prf  It is well-known and straightforward to verify that,
for any density matrix $\rho$ in $M_m$,
$\lim_{p \raw 1} \frac{1}{p- 1}\big(1-  \tr \rho^p \big) = S(\rho)$  and that 
$0 \leq S(\rho) \leq \log m$.     It then follows that \eqref{df1} holds; 
  the convergence is uniform in  $\gamma_{12}$ because the set of density 
  matrices  is compact.     By the mean value theorem, for any fixed $p, \gamma_{12}$
  one can find $\wtd{p}$ with $1 \leq \wtd{p} \leq p$ such that
$  u(p,\gamma_{12}) =   - \frac{d~}{dp} \,  
       \frac{ {\rm Tr}_{12} \, \gamma_{12}^{\wtd{p}}}{ {\rm Tr}_1 \, \gamma_1^{\wtd{p}}} \Big|_{p = \wtd{p}}$.
       Combining this with the fact that $ \gamma \geq \gamma^{\wtd{p}} \geq \gamma^2$ for
       any density matrix and $\wtd{p} \in (1,2]$ gives the following bound
\be 
|    u(p,\gamma_{12}) |   
       & = & \Big| \frac{ \trp_{12} \, \gamma_{12}^{\wtd{p}} \log \gamma_{12} \, \trp_1 \, \gamma_1^{\wtd{p}}
          -  \trp_{1} \, \gamma_{1}^{\wtd{p}} \log \gamma_{1} \,  \trp_{12} \, \gamma_{12}^{\wtd{p}} }
       {\trp_1 \, \gamma_1^{\wtd{p}}} \Big| \\
       & \leq &  \frac{S(\gamma_{12}) + S(\gamma_1) }{ {\rm Tr}_1 \, \gamma_1^2}  \qquad \qed
       \ee
       which is uniform  in $p$ for $p \in (1,2]$.

The quantity $S_{\rm cond}(\gamma_{12}) \equiv  S(\gamma_{12}) - S(\gamma_1)$ is called
the conditional entropy.  
Motivated by  \eqref{df1}, we   define the C.B. minimal entropy as
\be   \label{Scbmin}
  S_{\cbm}(\Phi) = \inf_{\psi \in \cd \ot \cd}  ~S_{\rm cond}\Big[ (\id \ot \Phi)\big( \proj{\psi} \big) \Big]
\ee 
and observe that it satisfies the following.
\begin{thm}    \label{thm:CB2}
For any CPT map $\Phi$,
\be
S_{\cbm}(\Phi) & = & \lim_{p \raw 1+}  \frac{1- \big[ \omega_p(\Phi) \big]^p}{p-1} 
\ee
where $\omega_p(\Phi)$ is given by \eqref{cb1p}.
\end{thm}
\prf  With $\gamma_{12} = (\id \ot \Phi) \big( \proj{\psi} \big)$, one finds
\be
 S_{\cbm}(\Phi) & = &  \inf_{|\psi \ket \in \cd \ot \cd} u(1,\gamma_{12} ) = 
   \inf_{\psi }  \, \lim_{p \raw 1+} \, u(p,\gamma_{12} )  \nn \\ \nn 
   & = &  \inf_{\psi }  \, \lim_{p \raw 1+} 
       \frac{1}{p- 1} \Big( 1 - 
     \frac{ {\rm Tr}_{12} \, \gamma_{12}^p}{ {\rm Tr}_1 \, \gamma_1^p}  \Big) \\
     & = &     \lim_{p \raw 1+}   \inf_{\psi}  \frac{1}{p-1} 
       \bigg( 1 -\frac{ \trp_{12} \gamma_{12}^p}{ \tr_1 \gamma_1^p } \bigg) \\  \nn 
     & =&    \lim_{p \raw 1+}    \frac{1}{p-1} 
       \bigg( 1 -  \sup_{\psi}  \frac{ \trp_{12} \gamma_{12}^p}{ \tr_1 \gamma_1^p } \bigg)  
        \ee
  where the interchange of $ \lim_{p \raw 1+}  $ and $   \inf_{\psi} $ is permitted
  by the uniformity in $\gamma_{12}$ of the continuity of $u(p,\gamma_{12}) $ at 
  $p = 1$. 
        
\begin{thm} For all  pairs of CPT maps $\Phi_A,  \Phi_B$,
\bee  S_{\cbm}(\Phi_A \ot \Phi_B)  = S_{\cbm}(\Phi_A ) + S_{\cbm}( \Phi_B)    \eee
\end{thm}
\prf   The result follows easily from the observations above and \eqref{mult.omega}.
\be
S_{\cbm}(\Phi_A \ot \Phi_B)  & = & \lim_{p \raw 1+}
     \frac{1 - \big[\omega_p(\Phi_A \ot \Phi_B)\big]^p }{p-1}  \\ \nn 
     & = &  \lim_{p \raw 1+}
     \frac{1 - \big[\omega_p(\Phi_A )\big]^p \, \big[ \omega(\Phi_B)\big]^p }{p-1} \\
     & = & \lim_{p \raw 1+} \frac{1 -  \big[ \omega_p(\Phi_A ) \big]^p }{p-1}   +   
      \bigg( \lim_{p \raw 1+}
        \big[ \omega_p(\Phi_A)\big]^p   \bigg) 
                  \lim_{p \raw 1+}  \frac{1 - \big[ \omega_p(\Phi_B)\big]^p  }{p-1}  \quad \nn \\ \nn
     & = &    S_{\cbm}(\Phi_A ) + S_{\cbm}( \Phi_B) 
\ee
where we used $ \lim_{p \raw 1+}   \big[ \omega_p(\Phi_A)\big]^p = 1$.     \qed

This result relies on \eqref{mult.omega} which is a special case of 
Theorem~\ref{thm:multqp} with $q =1$.   Recently, Jencova \cite{J}
found a simple direct proof of \eqref{mult.omega}.

 \subsection{Proof of CB additivity from SSA}  \label{sect:igor}

Recall that any CPT map $\Phi$ can be represented in the form
\be   \label{envrep}
   \Phi(\rho) = \trp_E U_{AE} \, \rho \ot \tau_E \, U_{AE}^{\dag}
\ee
with $U_{AE}$ unitary and $\tau_E$ a pure reference state on the 
environment.   The following key result follows from
standard purification arguments (which are summarized in Appendix~\ref{app}).
\begin{lemma}   \label{lemm:igor}
Let the CPT map $\Phi$ have a representation 
as in \eqref{envrep}.   One can find a reference system
$R$ and a pure state  $\proj{ \psi_{RA}}$ such that 
 $\trp_R \proj{ \psi_{RA}} = \rho$.   Define 
 $\gamma_{REA} = (I_R \ot U_{AE}) \, \big( \proj{ \psi_{RA}} 
\ot \tau_E \big) (I_R \ot U_{AE})^{\dag}$.  Then  $\gamma_{REA}$
is also pure and
\be
   S(\gamma_{EA}) - S(\gamma_E) = S(\gamma_{R}) - S(\gamma_{RA})
\ee
where the reduced density matrices are defined via partial traces.
\end{lemma}

It follows from \eqref{SSA} that the conditional entropy is subadditive, i.e., 
for any state $\gamma_{E_1E_2 A_1A_2}$,
\be  \label{condsub}
S(\gamma_{E_1E_2 A_1A_2}) - S(\gamma_{E_1E_2 })  & \leq &
  S(\gamma_{E_1  A_1 }) - S(\gamma_{E_1  }) +
    S(\gamma_{E_2 A_2}) - S(\gamma_{E_2 })
\ee
This was proved by Nielsen \cite{N} and appears as Theorem 11.16 in \cite{NC}.
It follows easily from the observation that \eqref{condsub} is the sum of the
following pair of inequalities, which are special cases of SSA
\bee
 S(\gamma_{E_1E_2 A_1A_2}) +  S(\gamma_{E_1 })  &  \leq &
  S(\gamma_{E_1  A_1 }) + S(\gamma_{E_1E_2 A_2  })  \\
   S(\gamma_{E_1E_2  A_2}) +  S(\gamma_{E_2 })  &  \leq &
  S(\gamma_{E_1  E_2 }) + S(\gamma_{E_2 A_2  }).
\eee
Now define
\be  \label{cbmin}
    S_{\cbm}(\Phi) = \inf_{\psi} \Big(  S \big[ (\id \ot \Phi)\big( \proj{\psi} \big)\big] - 
       S\big[ \trp_A  \proj{\psi} \big]  \Big),
\ee
Let $\Psi_{R A_1A_2}$ denote the minimizer for $\Phi_1 \ot \Phi_2$ and
\be
\gamma_{R_1R_2 A_1A_2E_1E_2} = \linebreak (I_R \ot U_{A_1E_1A_2E_2}) 
 \big( \proj{ \psi_{R A_1A_2}} 
\ot \tau_{E_1E_2} \big)(I_R \ot U_{A_1E_1A_2E_2})^{\dag}.
\ee
   Then
\begin{align}
S_{\cbm}(\Phi_1 \ot \Phi_2)  & =   
   S(\gamma_{R_1R_2 A_1A_2}) - S(\gamma_{R_1R_2 })  \nn \\
   & =      S(\gamma_{E_1E_2 })   - S(\gamma_{E_1E_2 A_1A_2})  \\ \nn
     & \geq   S(\gamma_{E_1  }) - S(\gamma_{E_1  A_1 })   +
 S(\gamma_{E_2 }) -   S(\gamma_{E_2 A_2}) . 
\intertext{Next, use the lemma to find  purifications  $\psi_{RA}^{\prime}$ and
$\psi_{RA}^{\prime \prime}$ so that the last line above} 
  & =  S(\gamma_{R_1 A_1 }^{\prime }) - S(\gamma_{R_1 }^{\prime })   +
 S(\gamma_{R_2 A_2 }^{\prime \prime}) -   S(\gamma_{R_2}^{\prime \prime})  \\ \nn
&  \geq   S_{\cbm}(\Phi_1 ) + S_{\cbm}( \Phi_2).    
\end{align}
The reverse inequality can be obtained using  product $\Psi$.

 \section{Completely bounded norms} \label{sect:CB}
 
 \subsection{Definitions }  \label{sect:CBdefs}

For the applications in this paper, we can define the completely bounded (CB) norm
of a map $\Phi :  L_q(M_m) \mapsto  L_p(M_n)$ as
\be   \label{CBdef1}
        \norm{ \Phi}_{\cbqp} \equiv \sup_d  \norm{ \id_d \ot \Phi}_{(\infty,q)  \raw (\infty,p) }
            = \sup_d  \bigg( \sup_Y \frac{ \norm{ (\id_d \ot \Phi)(Y) }_{(\infty,p)} }{ \norm{Y}_{(\infty,q)} } \bigg).
\ee
with 
 \be  \label{fm2}
\norm{Y}_{(\infty,p)} \equiv    \norm{Y}_{L_\infty(M_d;L_p(M_n))}  
   = \sup_{A,B \in M_d}   \frac{\norm{ (A \ot \rmi_n)Y (B\ot \rmi_n)}_p}
  {\norm{ A}_{2p} \, \norm {B}_{2p} } .
\ee
  Effros and Ruan  \cite{ER1,ER} introduced the
  norm  $\norm{Y}_{(1,p)}$.    Pisier \cite{PisOH,PisLp} subsequently
 used   complex interpolation between  them
 to define  a norm  $\norm{Y}_{(t,p)}$ for any $1 < t  < \infty$.
He   showed (Theorem 1.5 in \cite{PisLp}) that  the norm obtained
by this procedure satisfies
\be \label{fm3}
\norm{Y}_{(t,p)} \equiv  \nrm Y\|_{L_t(M_d;L_p(M_n))}  = 
   \inf_{ \substack{  Y=(A\ot \rmi_n)Z(B \ot \rmi_n) \\ A,B \in M_d }  }
        \nrm A\|_{2t} \, \nrm B\|_{2t}  \, \nrm Z\|_{(\infty,p)} ,  
\ee
which we can  regard as its definition.   The vector space $M_d \ot M_n$ equipped with the norm \eqref{fm3} is a Banach
space which we denote  by $L_t(M_d;L_p(M_n))$.   Given an operator 
$\Omega: L_t(M_d;L_q(M_m)) \mapsto L_s(M_{d'};L_p(M_n))$,   the usual norm
for linear maps from one Banach space to another becomes 
\be
   \norm{ \Omega} \equiv  \norm{ \Omega}_{(t,q) \raw (s,p)} = \sup_{ Q \in M_d \ot M_m}
       \frac{ \norm { \Omega(Q) }_{(s,p)} } {\norm {Q}_{(t,q)}}.
\ee
Theorem 1.5 and Lemma 1.7 in Pisier  \cite{PisLp} show that one can use this norm
to obtain another  expression for the CB norm
 \be   \label{CBdeft}
        \norm{ \Phi}_{\cbqp} \equiv \sup_d  \norm{ \id_d \ot \Phi}_{(t,q)  \raw (t,p) }
            = \sup_d  \bigg( \sup_Y \frac{ \norm{ (\id_d \ot \Phi)(Y) }_{(t,p)} }{ \norm{Y}_{(t,q)} } \bigg) 
\ee
valid for all $t \geq 1$.  In effect, we can replace   $\infty$   in \eqref{CBdef1}
by any $t \geq 1$.
In working with the CB norm, we will find it convenient to choose 
$t = q$ when $q \leq p$ and $t = p$ when $q \geq p$.
Thus our working definition of the CB norm is \eqref{CBdeft} with $t = \min \{q,p \}$.
For the applications considered in  Sections~\ref{sect:addcb} and \ref{sect:appl},
this becomes $t = q = 1$.

\rmk When $X > 0$, H\"older's inequality implies
\bee
  \norm{A X B^{\dag}}_p \leq \sqrt{ \norm{A X A^{\dag}}_p \, \norm{B X B^{\dag}}_p } \leq 
   \max \{ \norm{A X A^{\dag}}_p \, ,   \, \norm{B X B^{\dag}}_p \}
\eee
and the unitary invariance of the norm implies that 
$\norm{A X A^{\dag}}_p = \norm{ \, |A| \, X\,  |A| \, }_p $.  Therefore, when $X \geq 0$,
we can replace any expression of the form
$\sup_{A,B} \norm{A X B^{\dag}}_p$ by
$\sup_{A > 0}  \norm{A X A}_p$ irrespective of what other restrictions
may be placed upon $A,B$.        We will show that for
CP maps, the CB norm is unchanged if the supremum is taken over 
$Y > 0$.   (See Section~\ref{sect:3.2}, and Theorem~\ref{thm:qppos} and 
Corollary~\ref{cor:qppos} in Section~\ref{sect:CBmult}.)
Thus, when working with CP maps, one can generally assume
that $A = B > 0$ in expressions for $\norm{Y}_{(q,p)}$.

When $Y > 0$  combining \eqref{fm2} and \eqref{fm3}
gives the identity,
\be \label{pmaxmin}
  \norm{Y}_{(p,p)} =   \inf_{\substack{B > 0 \\ \tr B = 1}}  ~ \sup_{ \substack{A > 0 \\ \tr A = 1}} ~
      \norm{ (A \ot \rmi_n)^{\frac{1}{2p}} (B \ot \rmi_n)^{- \frac{1}{2p}} Y (B \ot \rmi_n)^{-\frac{1}{2p}}
       (A \ot \rmi_n)^{\frac{1}{2p}} }_p
\ee
for all $p \geq 1$.    Since Theorem~\ref{thm:fub}  implies that 
$ \norm{Y}_{(p,p)} = \norm{Y}_p$, this gives a variational expression
for the usual $p$-norm on $M_{dn} \simeq  M_d \ot M_n$.  The choice
$n = 1$ yields a max-min principle for the  $p$-norm on $M_d$.

The Banach space $ L_t(M_d;L_p(M_n))$ is a special case of
a more general  Banach space $ L_t(M_d;E)$  
 for which a norm is defined on $d \times d$ matrices with entries in
 an operator space $E$ as described in Section~\ref{sect:op.space}.  
 Because we use here only operators
 $\Phi:  L_q(M_m)   \mapsto L_p(M_n)$ rather than the general situation
 of operators $\Omega: E \mapsto F$ between Banach spaces $E,F$,
 we give explicit expressions only for norms on $ L_t(M_d;L_p(M_n))$.
On a few occasions we  need to consider spaces  
$L_t(M_d;E)$ with $E =   L_q(M_m;L_p(M_n))$; we  denote
the norm on these space by $\norm{Y}_{(t,q,p)}$.   In general
we will only encounter triples with two distinct indices and
will not need additional expressions for these norms.
Such cases  as $\norm{Y}_{(q,q,p)}$   reduce 
to $ L_q(M_{dm};L_p(M_n)) $ via the isomorphism between
$M_{dm} \ot M_n \simeq M_d \ot M_m \ot M_n$;  most
situations require only comparisons via Minkowski type inequalities
given in Section~\ref{sect:fubmink}.   In
section~\ref{sect:more} we show that $\norm{Y}_{(1,p,1)} = \norm{Y}_{(1,p)}$; this is
 needed only for the application in Section~\ref{sect:ent}.

\subsection{An important lemma}  \label{sect:3.2}  

We   illustrate the use of \eqref{fm3} by proving the following lemma,
which is a special case of a more general result in \cite{JZ}.
It plays a  key role in the multiplicativity results of Section~\ref{sect:q>p} for
$q \geq p$.   Although not needed for our main result, it also has important 
implications when $q \leq p$.   We first define  
 $\ds{ \nrm \Phi \nrm_{q \raw p}^+  = \sup_{ Q > 0} \frac{ \norm{\Phi(Q)}_p}{ \norm{Q}_q}}$.
   \begin{lemma}  \label{key}
Let $\Phi : L_q(M_m) \mapsto L_p(M_n)$ be a  CP map.  Then for every $r \geq 1$
the map  $ \Phi\ot \id_d : L_q(M_m; L_r(M_d)) \mapsto L_p(M_n; L_r(M_d))$ satisfies
\be   \label{genqtpt}
\nrm \Phi \ot \id_d \|_{(q,r) \raw (p,r)}  \leq \nrm \Phi\| _{q \raw p}^+ \ee
\end{lemma} 
\pf{Lemma}  
For any $Q$  \eqref{fm3} implies that   one can find $A, Y$ such that  
$Q = (A \ot \rmi) Y (B \ot \rmi)$  
and $ \norm{ Q }_{(q,r)} = \norm{ A}_{2q} \, \norm{ B}_{2q}  \, \nrm Y \| _{(\infty,r)}$.
Since  $\Phi$ is completely positive, one can find
$K_j$ satisfying \eqref{kraus}.    Let $V_A$ denote the
block row   vector with elements  \linebreak
$\big(K_1 A \ot \rmi_d, \, K_2 A \ot \rmi_d, \, \ldots , K_m A \ot \rmi_d \big)$,
and similarly for $B$.  Then
\be
   ( \Phi\ot \id_d)(Q) = V_A (\rmi_{\nu} \ot Y) V_B^{\dag} = 
       \sum_j   (K_j A \ot \rmi_d) Y  (B K_j^{\dag} \ot \rmi_d). 
\ee
(Note that  $\rmi_{\nu} \ot Y $ denotes a block diagonal matrix with $Y$ along
the diagonal 
with $\rmi_{\nu}$  the identity in an additional reference space used to
implement the representation \eqref{kraus}.   $Y$ itself is in the   tensor
product space $M_m \ot M_d $ on which $ \Phi\ot \id_d$ acts;  $K_j$ and
$A$  are in $M_m$.    We can extend  $V$ to an element of 
$M_{\nu} \ot M_m \ot M_d $ by adding rows of zero blocks; 
i.e., to $\sum_{i,j =1}^{\nu}    \delta_{i1} |i\kb j| \ot K_j A \ot \rmi_d $.)
Therefore, applying \eqref{fm3}  on this extended space gives
 \be 
  \nrm (\Phi\ot \id_d)(Q)\|_{(p,r)}
  & \leq & \nrm V_A^{\dag} V_A \|_{p}^{1/2} \,  \nrm V_B^{\dag} V_B \|_{p}^{1/2}
     \nrm \rmi_{\nu} \ot Y  \| _{(\infty, \infty,r)}  \\
  & = &    \bnorm{ \sum_j K_j^{\dag} A^{\dag} A K_j}_p  \, 
  \bnorm{ \sum_j K_j^{\dag} B^{\dag} B K_j}_p^{1/2} \, \nrm Y\|_{(\infty,r)}   \nn \\
&  =  &\nrm \Phi(A^{\dag} A)  \|_p^{1/2}   \, \nrm \Phi(A^{\dag} A)  \|_p^{1/2} \, \nrm Y\|_{(\infty,r)}   \\
 &   \leq  &  \norm{ \Phi}_{q \raw p}^+  \, \norm{ A^{\dag} A}_{q}^{1/2}  
    \, \norm{ B^{\dag} B}_{q}^{1/2} \, \nrm Y \|_{(\infty,r)}  \nn  \\ \nn 
& =   &  \norm{ \Phi}_{q \raw p}^+  \, \norm{ Q }_{(q,r)}  
 \ee
where we used $\norm{ A^{\dag} A}_{q}  =  \big( \norm{A }_{2q}\big)^2$.  \qed

The following corollary implies that for any $p,q$, the norm $ \nrm \Phi \nrm_{q \raw p}$
 is achieved  on a positive semi-definite matrix $Q > 0$.    This was proved earlier by
 Watrous  \cite{Wat}, resolving a question raised in \cite {KR3}.  In Section~\ref{sect:CBmult},
 we will see that  a similar result holds for  CB norms of CP maps.   This is stated as
 Theorem~\ref{thm:qppos} for $q \leq p$ and Corollary~\ref{cor:qppos} for $q \geq p$.
 \begin{cor}  \label{cor:pos} 
  Let $\Phi : L_q(M_m) \mapsto L_p(M_n)$ be a  CP map.  Then for all $q,p \geq 1$,
  the norm $\norm{ \Phi}_{q \raw p} = \norm{ \Phi}_{q \raw p}^+$  
\end{cor}
\prf  The choice $d = 1$ in Lemma~\ref{key} gives 
  $\norm{ \Phi}_{q \raw p} \leq \norm{ \Phi}_{q \raw p}^+$.   Since the reverse
  inequality always holds, the result follows.   
  
Note that one can similarly conclude that 
$\sup_d \norm{  \Phi \ot \id_d }_{(q,t) \raw (p,t)} =  \norm{ \Phi}_{q \raw p}^+$ so
that nothing would be gained by defining an alternative to the CB norm in
this way.    In Section~\ref{sect:appl} we show that the depolarizing channel gives
an explicit example of a map with $\norm{\Phi}_{\cb1p} > \norm{\Phi}_{1 \raw p}$.
It is worth commenting on the difference between this result and the proof by Amosov, Holevo and Werner \cite{AHW} that
$\norm{\id \ot \Phi}_{(1,1) \raw (p,p)} = \norm{ \Phi}_{1 \raw p}$.    In the latter,
the identity is viewed as an isometry from one Banach space $L_q(M_d)$
to another, $ L_p(M_d)$.   In the case of the CB norm, the identity is viewed
 as a map from the Banach space $ L_t(M_d)$ onto itself.   
 Thus,   we consider $\id_d \ot \Phi$ with 
 $\id_d: L_t(M_d) \raw L_t(M_d)$ and $\Phi : L_q(M_m) \mapsto L_p(M_n)$,
 for which we need to consider
 what  norm should be used on the domain $M_d \ot M_m$ if $t \neq q$ or on
 the range $M_d \ot M_n$ if $t \neq p$?  When $q \neq p$ this question is unavoidable.  
 One needs a norm  which acts like $L_t$ on $M_d$ and $L_p$ on $M_n$, and
 \eqref{fm2} provides such a norm.
Some of the motivation for the definitions used here is sketched in the next
section.

For a discussion of the stability properties of $\norm{\id \ot \Phi}_{(q,q) \raw (p,p)}$ 
see Kitaev \cite{Kv} and Watrous \cite{Wat}.  Note that in the case $q = p$, the
two types of norms for the extension $ \Phi \ot \id_d $ coincide and our results
imply that for CP maps 
$\norm{ \Phi}_{\rm CB, p \raw p} = \norm{  \Phi \ot \id_d }_{(p,p) \raw (p,p)} = 
     \norm{ \Phi}_{p \raw p}^+$.
     However, for measuring the difference between channels  \cite{Kv,Wat}, one is primarily
     interested in maps of the form $\Phi_1 -  \Phi_2 $   which are not CP.
  
\subsection{Operator spaces}  \label{sect:op.space}

The Banach space $E = L_p(M_n)$ together with
the sequence of norms   on the spaces $ L_\infty(M_d;L_p(M_n))$ with 
$d = 1, 2, \ldots$ form what is known as an {\em operator space}.
     More generally, an operator space is a Banach
space $E$ and a sequence of norms defined on the spaces $M_d(E)$,
whose elements are $d \times d$  matrices with elements in $E$,
with certain properties that guarantee that $E$ can be embedded in
${\cal B}({\cal H})$, the bounded operators on some Hilbert space
${\cal H}$.   Alternatively, one can begin with a  subspace 
$E \subset {\cal B}({\cal H})$; then the norm in $M_d(E)$ is
given by the inclusion  
$M_d(E) \subset M_d \big({\cal B}({\cal H}) \big) \simeq {\cal B}({\cal H}^{\ot d})$
consistent with interpreting an element of $M_d(E)$ as a block matrix.
(Usually such a situation is considered a concrete operator space in contrast
to an abstract operator space given by   matrix norms satisfying
Ruan's axioms \cite{ER,Pis3,Ruan}.)
The only operator  spaces we  use  in this paper
are those with $E = L_p(M_n)$ and, occasionally,  $E = L_t(M_d;L_p(M_n))$.
Although a concrete representation for even these spaces is not known,
the explicit expressions for the norms given in Sections~\ref{sect:CBdefs} 
and \ref{sect:more} suffice for many purposes.  (The reader who wishes to explore 
the literature should be aware that most of it is written in terms of
$ L_t(M_d;E)$ rather than $ L_t(M_d;L_p(M_n))$ and that the notation
$ S_t(M_d;E)$ (for Schatten norm) is more common than $L_t$.)

For   maps 
from ${\cal B}({\cal H})$ to ${\cal B}({\cal K})$ complete boundedness
is just uniform boundedness for the  sequence of norms of $\id_d \ot \Phi$ .
This notion is built in a manner  analogous to the familiar notion
of complete positivity.  In a similar way, one can define other ``complete''
notions, such as complete isometry 
based on the behavior of $\id_d \ot \Phi$.

The particular type of operator space considered here is called a 
``vector-valued $L_p$ space''.     We have already remarked on the need to
define a norm on $ L_t(M_d;L_p(M_n))$ to give a non-commutative 
generalization of the classical Banach space $\ell_t(\ell_p)$.
  Unfortunately, such naive generalizations as 
  $    \big( \sum_{jk}  \norm{ Y_{jk}}_p^t \big)^{1/t} $ or
  $   \big(  \trp_1 \big( \trp_2  |Y|^p \big)^{t/p} \big)^{1/t} $ do not even
  define norms.    The norms described in Section~\ref{sect:CBdefs}, although
  difficult to work with, yield an elegant structure with the following
  properties.
\begin{enumerate}

 \renewcommand{\labelenumi}{\theenumi}
    \renewcommand{\theenumi}{\alph{enumi})}
 
 \item  for the subalgebra of diagonal matrices the norm on $ L_t(M_d;L_p(M_n))$
    reduces to that on $\ell_t(\ell_q)$.
 
 \item  When $ Y = A \ot B$ is a tensor product,  ~  $\norm{ Y}_{t,p} = \norm{A}_t  \, \norm{B}_p
     = \big( \tr |A|^t \big)^{1/t} \, \big( \tr |B|^p \big)^{1/p}$.
  
  \item   The Banach space duality between $L_p$ and $L_{p^\prime}$ with
  $\tfrac{1}{p} + \tfrac{1}{p^{\prime}} = 1$ generalizes to
  \be
       L_q(M_d;L_p(M_n))^* =  L_{q^{\prime}}(M_d;L_{p^{\prime}}(M_n)).
  \ee
  
   \item   The collection of norms on $\{ L_t(M_d;L_p(M_n)) \}$ can be obtained
   from some (abstract) embedding of $L_p(M_d)$ into ${\cal B}({\cal H})$
   providing the operator space structure of $L_p(M_d)$.

     \item  The structure of $ L_t(M_d;L_p(M_n))$ can be used to develop a theory
  of vector-valued non-commutative  integration which generalizes
the theory of non-commutative integration developed by Segal
\cite{Segal} and Nelson \cite{Nelson}. 
   
 \end{enumerate}
 Although not used explicitly, properties (c) and (e) play an important role 
 in our results.    Consequences of (e) described in Section~\ref{sect:fubmink}
 play a key role in the proofs in Section~\ref{sect:CBmult}  and Section~\ref{sect:ent}.
 Theorem~\ref{thm:omeg1p}, which gives the simple expression \eqref{cb1p} 
 for the CB norm in  the case $1 \raw p$,  
 is an immediate consequence of a fundamental duality theorem.
 
For general information on operator spaces,
see Paulsen \cite{Paul},  Effros and Ruan \cite{ER} or Pisier \cite{Pis3}.  
The theory of non-commutative vector valued $L_p$ spaces  was developed
 by  Pisier in two monographs  \cite{PisOH} and \cite{PisLp}.  
 Additional developments can be found in \cite{Junge1} and \cite{Pis3}.
 
 \subsection{Fubini and Minkowski generalizations} \label{sect:fubmink}

Because vector valued $L_p$-spaces permit the
development of a consistent theory of vector-valued non-commutative 
integration, one would expect  
 generalizations of fundamental integration theorems.  This
 is indeed the case, and   analogues  of both Fubini's theorem
 and Minkowski's inequality play an important role in the results 
 that follow.

   First, Theorem 1.9 in \cite{PisLp} gives a non-commutative
   version of Fubini's theorem.
 \begin{thm}    \label{thm:fub}
For any  $1 \leq p \leq \infty$, the isomorphisms 
$L_p(M_d;L_p(M_n))  \simeq  L_p(M_d \ot M_n) \simeq L_p(M_{dn})$ 
hold in the  sense of complete isometry,  which implies that for all $W \in M_d \ot M_n$,
\be  \label{fub}
  \norm{W}_{L_p(M_d;L_p(M_n))} = \norm{W}_{L_p(M_n;L_p(M_d))} = 
     \norm{W}_p =   \big( \tr W^p \big)^{1/p} .
\ee
\end{thm}

The next result, which is Theorem 1.10 in \cite{PisLp},  will lead to
non-commutative versions of Minkowski's inequality and deals with
the flip map $F$ which takes $ A \ot B \mapsto B \ot A$ and is then extended
by linearity to arbitrary elements of a tensor product space so that
$W_{12} \mapsto W_{21}$.
\begin{thm}  \label{thm:mink}
 For $q \leq p$, the flip map 
 $ F: L_q(M_d;L_p(M_n)) \mapsto L_p(M_n;L_q(M_d))  $ is a complete contraction.
 \end{thm}
 The fact that $F$ is a contraction yields an analogue of Minkowski's
 inequality for matrices.
  \be   \label{mink}
      \norm{W_{21}}_{(p,q)}   = \norm{ F(W_{12})}_{(p,q)}   \leq 
         \norm{W_{12}}_{(q,p)}   \qquad   \text{for} \quad  q \leq p.
 \ee
The fact that $F$ is a {\em complete} contraction means that $\id \ot F$ is also
a contraction which yields a triple Minkowski inequality 
\be  \label{cb3mink}
    \norm{W_{132}}_{(q,p,q)} \leq   \norm{W_{123}}_{(q,q,p)}
\ee
when $q \leq p$.

 \rmk   To see why we regard \eqref{mink} as a non-commutative version
 of Minkowski's inequality, recall the usual $\ell_p(\ell_q)$ version.
   For $t \geq 1$,
  $      \Big[  \sum_j    \big( \sum_k |a_{jk}| \big)^t \Big]^{1/t}   \leq
                \sum_k    \Big( \sum_j |a_{jk}|^t  \Big)^{1/t}  $, and
Carlen and Lieb \cite{CL} extended this to positive semi-definite matrices 
\be  \label{mink.mat}
   \Big[ \trp_1 \Big( \trp_2 \, Q_{12} \Big)^t \Big]^{1/t}  \leq  
      \trp_2 \Big( \trp_1 \, Q_ {12}^t  \Big)^{1/t} 
\ee 
As in the case of the classical inequalities, \eqref{mink.mat}
holds for $t \geq 1$ and the reverse inequality holds for $t \leq 1$.  
Moreover, it follows that for  $R \geq 0$
\be  \label{mink.qp}
         \Big[ \trp_1 \Big( \trp_2 \, R_{12}^q \Big)^{p/q} \Big]^{1/p}  \leq  
    \Big[    \trp_2 \Big( \trp_1 \, R_{12}^p  \Big)^{q/p} \Big]^{1/q} 
      \qquad    \text{for} \quad  q \leq p.
\ee
 To see that  \eqref{mink.qp} and \eqref{mink.mat} are equivalent, let
 $t = p/q$, and $Q_{12} = R_{12}^p$.  Then raising both sides of \eqref{mink.qp}
 to the $q$-th power yields  \eqref{mink.mat}.
 
 In general, the quantity $  \big[ \trp_1 \big( \trp_2 \, R^p \big)^{q/p} \big]^{1/q}$
      does not define a norm.
 Carlen and Lieb \cite{CL} conjectured
 that   $ \trp_1 \big( \trp_2 \, R^p \big)^{1/p} $  does
 define a norm for $1 \leq p \leq 2$, but proved it only in the
 case $p = 2$.   (For $p > 2$ it can be shown not to be a norm.)
 Their conjecture is that
\be  \label{mink3}
  \trp_3 \Big[ \trp_2 \Big( \trp_1 \, Q_{123} \Big)^t \Big]^{1/t}  \leq  
      \trp_{1,3} \Big( \trp_2 \, Q_ {123}^t  \Big)^{1/t} 
\ee 
which is very similar in form to \eqref{cb3mink} with $q =1, p = t$.


\subsection{More facts about $L_q(M_d;L_p(M_n))$ norms}  \label{sect:more}

We now state  two additional formulas
for  norms on   $L_q(M_d;L_p(M_n))$.    Although not needed for
the main result, some consequences are needed for
Theorem~\ref{thm:qppos} and in Section~\ref{sect:ent}.
  For detailed proofs see \cite{Junge1}.

We state both under the assumption  $1\le q \le p\le \infty$ and  $\frac1q=\frac1p+\frac1r$. Then
\begin{align} 
\norm{Y}_{(p,q)}  ~  \equiv ~  \nrm Y\|_{L_p(M_d;L_q(M_n))}  & ~ =  ~ \sup_{A,B \in M_d} 
  \frac{\nrm (A \ot \rmi_n)Y (B\ot \rmi_n)\|_{q}}
  {\nrm A\|_{2r} \, \nrm B\|_{2r}}  \label{fm4} \\
\intertext{and} 
  \label{fm5}
 \norm{Y}_{(q,p)}   ~  \equiv ~  \nrm Y\|_{L_q(M_d;L_p(M_n))}  & ~ =  
      \inf_{ \substack{  Y=(A\ot \rmi_n)Z(B \ot \rmi_n) \\ A,B \in M_d }  }
       \nrm A\|_{2r} \, \nrm B\|_{2r}  \, \nrm Z\|_p  
\end{align}
Moreover, when $Y > 0$ is positive semi-definite, one can restrict both
optimizations to $A = B > 0$.   In the case $X > 0$, $q = 1$, \eqref{fm4} becomes
\be
    \norm{ X_{12} }_{(p,1)} & = & \sup_{A > 0} \frac{ \norm{ (A \ot \rmi_n) X_{12} (A \ot \rmi_n)}_1}
        { \norm{A}_{2p^{\prime}}^2}    \nn \\   \label{fm4b}
          & = & \sup_{A > 0} \frac{ \tr A^2 X_1 }{ \norm{A^2}_{p^{\prime}} } = \norm{X_1}_p
\ee
and \eqref{fm5} can be rewritten   as
\begin{align}  \label{fm5b}
   \norm{X}_{(1,p)}  & = \inf_{\substack{A > 0 \\ X  =(A\ot \rmi_n)Z(A \ot \rmi_n)}}    
       \norm{A}_{2p^{\prime}}^2  \, \norm{Z}_p \\ \nn 
       & =  \inf_{ B > 0 , \,  \norm{B}_1 = 1}  
          \norm{ (B^{-1/  2p^{\prime}} \ot \rmi_n) \, X \, (B^{-1/  2p^{\prime}} \ot \rmi_n) }_p \\  \label{fm5c}
       & =  \inf_{ B > 0 , \,  \norm{B}_1 = 1}    
          \norm{ (B^{- \haf(1 - \frac{1}{p})} \ot \rmi_n) \, X \, (B^{- \haf(1 - \frac{1}{p})} \ot \rmi_n) }_p
          \end{align}
          
          In Section~\ref{sect:ent}, we will also need
\be   \label{trip.1p1}
    \norm{W_{132}}_{(1,p,1)} & = &  \norm{W_{132}}_{L_1(M_d;L_p(M_n;L_1(M_m)))} \nn \\
      & = &  \inf_{ \substack{ A \in M_d, A > 0 \\ W_{132} = 
           (A  \ot \rmi_{ 32}) Z_{132}  (A  \ot \rmi_{32}) } }
        \norm{A}_{2p^{\prime}}^2 \, \norm{Z_{132}}_{(p,p,1)}  \\  \nn
        & = &  \inf_{B_1  > 0, \norm{B_1}_1}  
           \norm{ (B_1^{-1/{2p^{\prime}} }\ot \rmi_{3} \ot \rmi_{2}) W_{132}  
            (B_1^{-1/{2p^{\prime}} }\ot \rmi_{3} \ot \rmi_{2})}_{(p,p,1)} \\
            & = &  \inf_{B_1  > 0, \norm{B_1}_1}    \label{trip.1p1.13}
             \norm{ (B_1^{-1/{2p^{\prime}} }\ot \rmi_{3}  ) W_{13}  
            (B_1^{-1/{2p^{\prime}} }\ot \rmi_{3} )}_{(p,p)}   \\  \nn
            & = & \norm{W_{13}}_{(1,p)}     \label{1p1red}
\ee
where \eqref{trip.1p1} is proved in \cite{Junge1} and
the reductions which follow used  \eqref{fm4b} and \eqref{fm4}.

\begin{lemma}   \label{lemm:fm6}
When  $1\le q \le p\le \infty$ and $X$ is a contraction, then
\be   \label{fm6}
 \norm{ C^{\dag} X D}_{(q,p)} \leq  
     \big( \norm{C^{\dag} C}_{(q,p)} ~ \norm{D^{\dag} D}_{(q,p)} \big)^{1/2}
\ee
\end{lemma}
\prf  It follows from \eqref{fm5} that one can find $A,B \in M_d$ and $Y,Z \in M_{dn}$
such that  $A,B > 0$, $\norm{A}_{2r} = \norm{B}_{2r} = 1$, $Y,Z > 0$ and
\begin{align*} 
    C^{\dag} C & = (A \ot \rmi_n) Y (A \ot \rmi_n)  & 
        \norm{C^{\dag} C}_{(q,p)}  & =   \norm{(A \ot \rmi_n) Y (A \ot \rmi_n)}_p \\
         D^{\dag} D & = (B \ot \rmi_n) Z (B \ot \rmi_n)  & 
        \norm{D^{\dag} D}_{(q,p)}  & =   \norm{(B \ot \rmi_n) Z (B \ot \rmi_n)}_p ~.
\end{align*} 
Moreover, there are partial isometries, $V,W$ such that
$C = V Y^{1/2} (A \ot \rmi_n)$ and $D = W Z^{1/2} (B \ot \rmi_n)$.  Then
\be
  C^{\dag} X D  =  (A \ot \rmi_n) Y^{1/2} V^{\dag} X W Z^{1/2} (B \ot \rmi_n)
\ee
and it follows from \eqref{fm5} and H\"older's inequality that
\be
  \norm{ C^{\dag} X D}_{(q,p)} & \leq  & 
     \norm{ (A \ot \rmi_n) Y^{1/2} \, V^{\dag} X W \, Z^{1/2} (B \ot \rmi_n) }_p   \\
     & \leq &  \norm{ (A \ot \rmi_n) Y A \ot \rmi_n)}_p^{1/2} \,  \norm{ V^{\dag} X W }_{\infty}
        \,   \norm{ (B \ot \rmi_n) Z B \ot \rmi_n)}_p^{1/2}   \nn \\   \nn
         & = &    \norm{C^{\dag} C}_{(q,p)}^{1/2} ~ \norm{D^{\dag} D}_{(q,p)}^{1/2}
         \qquad  \qquad \qed
\ee

\subsection{State representative of a map}  \label{sect:rep}

A linear map $\Phi: M_d \mapsto M_d$ can be associated
with a  block matrix in which the $j,k$ block
is the matrix  
$\Phi \big( |e_j \kb e_k | \big)$ in the standard basis.   
This is often called
the ``Choi-Jamiolkowski matrix'' or ``state representative''
in quantum information theory and will be denoted $X_{\Phi}$.
Thus, 
\be   \label{choi}
  X_{\Phi} & = & \sum_{jk}  |e_j \kb e_k |  \ot \Phi \big( |e_j \kb e_k | \big) \ee    
  Choi \cite{Choi} showed that the map
$\Phi$ is CP if and only if  $X_{\Phi} $ is positive semi-definite.
Conversely given a (positive semi-definite) $d^2 \times d^2$
matrix $X$, one can use \eqref{choi} to define a  CP map $\Phi$.    In addition,
Choi showed that the eigenvectors of  $X_{\Phi}$ can be
rearranged to yield operators, $K_j$ such that
\be  \label{kraus}
    \Phi(Q) = \sum_j  K_j Q K_j^{\dag} .
\ee
This result  representation was obtained independently by Kraus \cite{Kr1,Kr2} 
and can be recovered from that of Stinespring \cite{Spg}.
 
For every CP map $\Phi$ with Choi matrix $X_{\Phi}$,  it follows from \eqref{choi} that
 \be \label{Xequiv}  
 \norm {(A  \ot \rmi) X_{\Phi} (A \ot \rmi)}_{p}    
    & = &  \norm{  \sum_{jk}  A |e_j \kb e_k |  A \ot \Phi \big(  |e_j \kb e_k |    \big) }_p\\
  & = &  \norm{ (\id \ot \Phi)\big( \proj{\psi_A} \big) }_p  \nn 
   \ee
where the last equality follows if we choose
$ |\psi_A \ket = \sum_j A | e_j \ket \ot  | e_j \ket   $

\begin{thm}    \label{thm:omeg1p}
For any CP map $\Phi$, 
{\em \be \label{sup.psi}
   \norm{ \Phi}_{\cb1p} = \norm{X_{\Phi}}_{(\infty,p)}
    =   \sup_{\norm{\psi} =1}   \frac{\norm{ (\id \ot \Phi)\big( \proj{\psi} \big) }_p}
                 { \norm{\trp_2 \,( \proj{\psi} )}_p}  
              ~     \equiv    ~ \omega_p(\Phi)  
 \ee}
\end{thm}
\prf  This result requires a fundamental duality result proved by
Blecher and Paulsen \cite{BP}  and by Effros and Ruan \cite{ER1,ER}
  and described in Section 2.3 of \cite{Pis3}.
It states that
\be
    \norm{ \Phi}_{\cb1p} = \norm{\Phi^*}_{{\rm CB}, p^{\prime} \raw \infty} = \norm{X_{\Phi}}_{(\infty,p)}
\ee
Using \eqref{fm2} gives
 \be
   \norm{ \Phi}_{\cb1p} & = &         
              \sup_{A  > 0 }  \frac{ \norm{ (A  \ot \rmi) X_{\Phi} (A \ot \rmi)}_p}   
             {  \norm{A^2}_p }   \nn  \\ \nn
              & = &  \sup_{ \psi }   \frac{\norm{ (\id \ot \Phi)\big( \proj{\psi} \big) }_p}
                 {   \norm{\trp_2 \,( \proj{\psi} )}_p} .  
          \ee
  Since the ratio is unchanged if $|\psi \ket$ is multiplied by a constant,
 one can restrict the supremum above to $\norm{\psi} =1$. \qed

\section{Multiplicativity for CB norms}  \label{sect:CBmult}

\subsection{ $1 \leq  q \leq p$} \label{sect:subq<p}

We now prove  multiplicativity of the CB norm for maps 
 $\Phi : L_q(M_m) \mapsto   L_p(M_m) $ with $q \leq p$.  
\begin{thm}   \label{thm:multqp} 
Let $q \leq p$ and $\Phi_A: L_q(M_{m_A}) \mapsto  L_p(M_{n_A})$
and  $\Phi_B: L_q(M_{m_B }) \mapsto  L_p(M_{n_B})$ be CP and CB. Then
\be \norm  {\Phi_A \ot  \Phi_B }_{\cbqp}  =  \nrm \Phi_A \|_{\cbqp} \nrm
 \Phi_B \|_{\cbqp} ~ .\ee
\end{thm}
\prf  
Let $Q_{CAB}$ be in $M_d \ot M_{m_A} \ot M_{m_B}$ and
$R_{CAB} = (\id_d \ot \id_{m_A} \ot \Phi_B)(Q_{CAB})$.  Then using \eqref{cb3mink},
one finds
\be
\lefteqn{  \norm{\Phi_A \ot \Phi_B}_{\cbqp} =  \sup_d \sup_{Q_{CAB}} 
   \frac{  \norm{ (\id_d \ot \Phi_A \ot \Phi_B)  Q_{CAB}}_{(q,p,p)}  }
      { \norm{ Q_{CAB}}_{(q,q,q)}  }  }  \\
  & ~ & \nn  \\ & = &   \sup_{Q_{CAB} }
   \frac{    \norm{ (\id_d \ot \Phi_A \ot \id_{n_B})  R_{CAB}}_{(q,p,p)}  }
   { \norm{ R_{CAB}}_{(q,q,p)} } ~
       \frac{ \norm{ (\id_d \ot \id_{m_A} \ot \Phi_B)(Q_{CAB})}_{(q,q,p)}  } 
     { \norm{ Q_{CAB}}_{(q,q,q)}  }   \nn \\
     & ~ & \nn \\   
& \leq &   \sup_{R_{CBA} }
   \frac{  {  \norm{ (\id_d  \ot \id_{n_B} \ot \Phi_A)  R_{CBA}}_{(q,p,p)}   }}
   {\norm{ R_{CBA}}_{(q,p,q)}}  ~ 
         \frac  {\norm{ R_{CBA}}_{(q,p,q)}} { \norm{ R_{CAB}}_{(q,q,p)}} \\
  & ~ &  \qquad \qquad \qquad \qquad  \qquad \times     
  \sup_{Q_{CAB}}   
      \frac{ \norm{ (\id_d \ot \id_{m_A} \ot \Phi_B)(Q_{CAB})}_{(q,q,p)}  } 
     { \norm{ Q_{CAB}}_{(q,q,q)}  }         \nn \\
     & \leq &  
       \norm{  \id_{n_B} \ot \Phi_A }_{{\rm CB},(p,q) \raw (p,p)}~ \norm{\Phi_B}_{\cbqp}   \\
   & = &      \norm{\Phi_A}_{\cbqp}    ~ \norm{\Phi_B}_{\cbqp}  .  \nn
   \ee 
For the last two lines, we
used $ \norm{  \id_n \ot \Phi_A }_{{\rm CB},(p,q) \raw (p,p)}$ to denote the CB norm of  
$ \id_n \ot \Phi_A :  L_p(M_n;L_q(M_m)) \mapsto L_p(M_n;L_p(M_m)) $ and then
applied Corollary 1.2 in \cite{PisLp}, which states that this is the same as the
CB norm of $\Phi : L_q(M_m) \mapsto L_p(M_m) $.  

 To prove the reverse direction,  we need a slight modification of the
standard strategy of showing that the bound can be achieved with a
tensor product.  It can happen that the CB norm itself is not attained
for any finite $I_d \ot \Phi $ norm.     Therefore, we first show that any finite
product can be achieved, and then use the fact that the CB norm can be
approximated arbitrarily closely by such a product.

Thus, we begin with the observation that for any  
$d$ and $X,Y$ in the unit balls for  $ L_q(M_d \ot M_m)$ and $L_q(M_d \ot M_n)$,  
 there exist $Q, R > 0$ in the unit ball of $L_{2q}(M_d)$ such that
\be \nrm (Q \ot 1_m)[ \id_d \ot \Phi_A(X )](Q \ot  \rmi_m)\|_{q} 
 =  \nrm ( \id_d \ot  \Phi_A(X))\|_{L_q(M_d;L_p(M_m))} 
 \ee
and
 \be
 \nrm (R \ot  \rmi_n)( \id_d \ot \Phi_B(Y  ))(R \ot  \rmi_n)\|_{q} 
    =  \nrm [ \id \ot  \Phi_B(Y  )]\|_{L_q(M_d;L_p(M_n))} ~ .
 \ee
Then,   using
Theorem~\ref{thm:fub}, one finds
 \be 
\lefteqn{ \nrm \Phi_A\ot \Phi_B\|_{\cbqp}  ~ \geq ~ 
   \nrm [\id_{M_{d^2}} \ot (\Phi_A \ot \Phi_B)]  (X \ot Y  )\|_{L_q(M_{d^2};L_p(M_{mn}))} }
   \qquad  \nn \\
  &  \geq & \nrm (Q \ot R \ot \rmi_{mn} )[ \id_{{d^2}} \ot (\Phi_A\ot \Phi_B)](X \ot  Y  )
    (Q \ot R \ot \rmi_{mn})\|_{q}  \nn  \\
  &  = & \nrm (Q \ot \rmi)[\id_{M_d}\ot \Phi_A(X )](Q\ot \rmi)\|_{q} ~
  \nrm  (R\ot \rmi)[ \id_{M_d}\ot \Phi_B(Y  )](R\ot \rmi) \|_{q}  \nn \\
  &  = &  \nrm ( \id_{M_d} \ot \Phi_A(X ))\|_{L_q(M_d;L_p(M_m))} ~ 
   \nrm ( \id_{M_d} \ot  \Phi_B(X ))\|_{L_q(M_d;L_p(M_n))}    \label{lastcbqp}
 \ee 
Given $\eps > 0$, one can find   $d,X ,Y  $ such  that 
$  \nrm \Phi_A\|_{\cbqp}  <  \eps + \nrm ( \id_{M_d} \ot \Phi_A(X ))\|_{L_q(M_d;L_p(M_m))}$
and   
$  \nrm \Phi_B\|_{\cbqp}  <  \eps + \nrm ( \id_{M_d} \ot \Phi_B(X ))\|_{L_q(M_d;L_p(M_n))}$.
Inserting this in \eqref{lastcbqp} above gives
\bee
\nrm \Phi_A\ot \Phi_B\|_{\cbqp} ~ \geq ~  \nrm \Phi_A\|_{\cbqp}  \nrm \Phi_B\|_{\cbqp} \,  -  \, 
       \eps\big(\nrm \Phi_A\|_{\cbqp} + \nrm \Phi_B\|_{\cbqp}\big) +  O(\eps^2) 
\eee 
Since $\eps>0$ is arbitrary, we can conclude that
 \bee
 \nrm \Phi_A\ot \Phi_B\|_{\cbqp} \geq  \nrm \Phi_A\|_{\cbqp} \nrm \Phi_B\|_{\cbqp}.   \qquad \qed
 \eee
 
 The next result implies that for CP maps, it suffices to restrict the
 supremum in the CB norm to positive semi-definite matrices.
\begin{thm}  \label{thm:qppos}
When $q \leq p$ and $\Phi : L_q(M_m) \mapsto L_p(M_n)$ is CP, 
$\norm{ \id_d \ot  \Phi}_{(q,q) \raw (q,p)}$ is achieved with a positive
semi-definite matrix, i.e.,  $\norm{ \id_d \ot  \Phi}_{(q,q) \raw (q,p)}
  =  \norm{ \id_d \ot  \Phi}_{(q,q) \raw (q,p)}^+  $.
\end{thm}
\prf First use the polar decomposition of $Q \in M_{dm}$ to write
 $Q =  Q_1^{\dag} Q_2$ with $Q_1 =   |Q|^{1/2} U, ~ Q_2 = |Q|^{1/2}$ where 
 $U$ is a partial isometry and
 $|Q| = (Q^{\dag} Q)^{1/2}$.   The matrix 
 \be
  \pmx Q_1^{\dag} \\   Q_2^{\dag} \emx \pmx Q_1 &   Q_2  \emx =
 \pmx Q_1^{\dag} Q_1   & Q_1^{\dag} Q_2  \\  Q_2^{\dag} Q_1 & Q_2^{\dag} Q_2 \emx 
  =     \pmx U^{\dag} |Q|U   & Q  \\  |Q| & Q^{\dag}   \emx 
  > 0
  \ee
 is positive semi-definite.  Since $\Phi$ is CP,  so is $\id \ot \Phi$ which implies that
\be   \label{p1}
  \pmx (\id \ot \Phi)(U^{\dag} |Q|U )   & (\id \ot \Phi)(Q)  \\  
     (\id \ot \Phi)(Q ^{\dag}  ) & (\id \ot \Phi)(  |Q|  ) \emx > 0
\ee 
is positive semi-definite.  We now use the fact that
a $2 \times 2$ block matrix $\pmx A & C \\ C^{\dag} & B \emx$ with $A,B > 0$ is
positive semi-definite if and only if $C = A^{1/2} X B^{1/2}$ with $X$ a contraction.
Applying this to \eqref{p1} gives
 \be      (\id \ot \Phi)(Q) = [(\id \ot \Phi)(U^{\dag} |Q|U )]^{1/2} \, X  \, [(\id \ot \Phi)(  |Q|   )]^{1/2} \ee
  with $X$ a contraction.  Therefore, it follows from \eqref{fm6} that
 \be
    \norm{ (\id \ot \Phi)(Q)}_{(q,p)} & = & 
        \norm{   [(\id \ot \Phi)(U^{\dag} |Q|U )]^{1/2}  \, X \,  [(\id \ot \Phi)(  |Q|   )]^{1/2} }_{(q,p)}   \\ \nn
        & \leq &   \Big( \norm{  (\id \ot \Phi)(U^{\dag} |Q|U )}_{(q,p)} \,
           \norm{ (\id \ot \Phi)(  |Q|   )}_{(q,p)}  \Big)^{1/2} \\  \nn
           & \leq &  \norm{\id \ot \Phi }_{(q,q) \raw (q,p)}^+  ~
               \big( \norm{ | Q | }_q  \, \norm{U^{\dag} |Q|U }_q \big)^{1/2} \\ \nn
               & = & \norm{\id \ot \Phi }_{(q,q) \raw (q,p)}^+  \, \norm{|Q|}_q   \qquad  \qed
         \ee

  \subsection{$q \geq p$} \label{sect:q>p}

\begin{thm}\label{thm:q>p} 
Let $q \ge p$ and $\Phi_A:L_q(M_{m_A})\to L_p(M_{n_A})$,
$\Phi_B:L_q(M_{m_B})\to L_p(M_{n_B})$
 be maps which are both CP. Then
\begin{align}  \label{q>pt1}
& \text{a)} &    \qquad  \qquad   &   
\quad  \nrm \Phi\|_{\cbqp}    =  \nrm \Phi \nrm_{q \raw p}  =   \nrm \Phi \nrm_{q \raw p}^+   & & \\  \label{q>pt2}
& \text{b)} &    \qquad  \qquad       &  
  ~~~~      \nrm \Phi_A \ot \Phi_B\|_{q \raw p}   =  \nrm \Phi_A\|_{q \raw p} \nrm  \Phi_B\|_{q \raw p} & & \\
   \label{q>pt3}
& \text{c)} &   \qquad  \qquad      &    \nrm \Phi_A\ot \Phi_B\|_{\cbqp}   =   \nrm \Phi_A\|_{\cbqp} \nrm  \Phi_B\|_{\cbqp}&  ~.
 &     \end{align}
\end{thm}
Combining part (a) with Corollary~\ref{cor:pos} implies that it suffices to restrict the
 supremum in the CB norm to positive semi-definite matrices.
\begin{cor}  \label{cor:qppos}
When $q \geq p$ and $\Phi : L_q(M_m) \mapsto L_p(M_n)$ is CP, 
$\norm{ \id_d \ot  \Phi}_{\cbqp}$ is achieved with a positive
semi-definite matrix.
\end{cor}

\noindent{\bf Proof of Theorem~\ref{thm:q>p}:}
To prove part (a), observe that
\be   \label{pfa1}
 \norm{\Phi}_{\cbqp} & = &  \sup_d   \bigg( \sup_{W_{AB}  \in M_d \ot M_m}
 \frac{  \norm{(\id_d \ot \Phi )(W_{AB})}_{p} }
    {\norm{W_{AB}}_{(p,q)} }  \bigg)  \nn \\    \label{pfa2}
    & =  &    \sup_d   \bigg( \sup_{W_{AB}}  \frac{   \norm{(\id_d \ot \Phi )(W_{AB})}_p } 
     {\norm{W_{BA}}_{(q,p)}} ~
   \frac{    \norm{W_{BA}}_{(q,p)} }   {\norm{W_{AB}}_{(p,q)} } \bigg) \\ \nn 
    \label{pfa3}       & \leq  &       
        \sup_d  \sup_{W_{BA}  \in M_m \ot M_d}  \frac{  \norm{(\Phi \ot \id_d )(W_{BA})}_p }   {\norm{W_{BA}}_{(q,p)} } \\
          & \leq &  \norm{\Phi}_{q \raw p}^+
\ee
The first inequality follows from the fact that the
second ratio in \eqref{pfa2} is $\leq 1$ by \eqref{mink} 
 and the last inequality then follows from \eqref{genqtpt}.
When $d = 1$, the 
supremum over $W$ of the ratio in \eqref{pfa1} is precisely 
$\norm{\Phi}_{q \raw p}$ which implies 
$\norm{\Phi}_{\cbqp}   \geq   \norm{\Phi}_{q \raw p}$.  
This proves part (a).

To prove part (b), write $\Phi_A \ot \Phi_B = (\Phi_A \ot \id )(\id \ot \Phi_B)$
and for any $Q_{AB}  \in M_{m_A} \ot M_{m_B}$, 
let $R_{AB} =  (\id \ot \Phi_B)(Q)$.  Then
\be   \label{pfb1}
\lefteqn{ \norm{ \Phi_A \ot \Phi_B}_{q \raw p} = 
    \sup_Q  \frac{ \norm{ (\Phi_A \ot \Phi_B)(Q)}_p} { \norm{Q}_q} } ~~ \\
    \label{pfb2}
    & \leq  & \sup_Q \, \frac{ \norm{(\Phi_A \ot \id )(R_{AB}) }_{p} }
       { \norm{R_{AB}}_{(q,p)} }  ~
       \frac{ \norm{R_{AB}}_{(q,p)} }{ \norm{R_{BA}}_{(p,q)} } ~
       \frac{ \norm{( \Phi_B \ot \id)(Q_{BA})}_{(p,q)} }{ \norm{Q }_{q}}  \nn  \\
   & \leq  &    \sup_R \, \frac{ \norm{(\Phi_A \ot \id )(R) }_{(p,p)}} 
       { \norm{R}_{(q,p)} } ~ \sup_{Q_{BA}} 
        \frac{ \norm{( \Phi_B \ot \id)(Q_{BA})}_{(p,q)} }{ \norm{Q_{BA}}_{q,q} }  \\ \nn 
    & \leq  &  \norm{ \Phi_A }_{q \raw p}   ~   \norm{ \Phi_B }_{q \raw p}  
\ee
where we used \eqref{mink}, Fubini,  and $R_{BA} =  (\Phi_B \ot \id)(Q_{BA})$.
This proves (b).

Part (c) then follows immediately from (a) and (b).  \qed



\section{Applications  of CB entropy}  \label{sect:appl}

\subsection{Examples and bounds}   \label{sect:exam}

It is well-known that conditional information can be
negative as well as positive.   Therefore, it is not surprising that
\eqref{cbmdef}  can also be either positive or negative, depending
on the channel $\Phi$.
As in Section~\ref{sect:intro}, we adopt the convention that
$\gamma_{12} = (\id \ot \Phi) \big(|\psi \kb \psi| \big) $.   One has 
the general bounds
\be   \label{bd1}
     - S(\gamma_1)   \leq S_{\cbm}(\Phi)   \leq S(\gamma_1) 
     \ee
 which imply
 \be  \label{bd2}
      - \log d & \leq S_{\cbm}(\Phi)   \leq \log d .
\ee
The lower bound in \eqref{bd2}  follows from the definition \eqref{cbm2}
and the positivity of the entropy $S(\gamma_{12})  > 0$;
the upper bound follows from subadditivity  
$S(\gamma_{12}) \leq S(\gamma_1) + S(\gamma_2)$.
The upper bound is attained if and only if the output
$(\id \ot \Phi) \big( \proj{ \psi} \big) $ is always a product.
The lower bound in \eqref{bd2} is attained for the identity channel,
and the upper bound for the completely noisy channel 
$\Phi(\rho) =  (\tr  \rho) \tfrac{1}{d} I$.

Next, consider the depolarizing channel  
 $\Omega_{\mu}(\rho) = \mu \rho + (1-\mu) (\tr \rho) \tfrac{1}{d} I$.
 This channel  satisfies the covariance condition 
 $U \Omega(\rho) U^* = \Omega(U \rho U^*)$ for all unitary $U$.  Lemma 2 in
 the appendix of \cite{J} can therefore be used to show that the minimal CB entropy
is achieved when $\gamma_1 = \trp_2 (\id \ot \Omega)(\proj{\psi})$ is
the maximally mixed state $\frac{1}{d} \rmi $ so that $|\psi\ket$ is maximally
entangled and
\be 
\norm{\Omega}_{\cb1p} = S(X_\Omega) - \log d\
\ee
Moreover, the state $(\id \ot \Omega_{\mu})(\proj{\psi})$ has  one non-degenerate
 eigenvalue $\tfrac{1 +(d^2 \! - \! 1) \mu}{d^2}$ and the eigenvalue 
 $\tfrac{1-\mu}{d^2} $
 with multiplicity $d^2 - 1$.  From this one finds
 \begin{align}   \label{dep.omegp}
    \omega_p(\Omega_{\mu}) & = d^{- \frac{p+1}{p}} \Big[  (1 - \mu + d^2 \mu)^p
        + (d^2 -1)(1 - \mu)^p \Big]^{1/p}
    \intertext{and}
     S_{\cbm}(\Omega_{\mu} ) & =    - \tfrac{1-\mu}{d^2}  \log \tfrac{1-\mu}{d^2} 
       - (d^2 -1) \tfrac{1-\mu}{d^2} \log \tfrac{1-\mu}{d^2} - \log d \\  \nn 
    & =   \log d - \tfrac{1}{d^2} \big[ (1\mm  \mu \pp  d^2 \mu) \log (1\mm  \mu \pp d^2 \mu)
       + (d^2 \mm 1) (1 \mm  \mu) \log (1 \mm  \mu)
 \end{align}
  In the case of qubits, $d = 2$ and \eqref{dep.omegp} becomes
  \begin{align}
  \norm{ \Omega_{\mu}}_{\cb1p} =  \omega_p(\Omega_{\mu}) & = 
         2^{-(p+1)/p} \big[  (1+3\mu )^p + 3(1-\mu )^p \big]^{1/p}
\intertext{         which can be compared to} 
      \norm{ \Omega_{\mu}}_{1 \raw p} =  \nu_p(\Omega_{\mu})  & =
        2^{-1 } \big[  (1+ \mu )^p +  (1-\mu )^p \big]^{1/p} .
  \end{align}
 The strict convexity of  $f(x) = x^p$ implies that
 for $\mu   > 0$,  
 \bee
 (1 + \mu )^p = \big( \tfrac{(1 + 3\mu ) + (1-\mu )}{2} \big)^p
     < \haf \big[ (1+3\mu )^p + 3(1-\mu )^p \big]
 \eee
  from which it follows that    
 $\norm{  \Omega_{\mu}}_{\cb1p} > \norm{ \Omega_{\mu}}_{1 \raw p}$.
    This confirms that, in general, the   CB norm $\norm{\Phi}_{\cb1p}$
    of a map $\Phi$ is strictly greater than
$\norm{ \Phi}_{1 \raw p}$.  (This   can be  seen directly  for the 
identity map $\id$ which corresponds to $\mu = 1$.)
 For qubits, one can verify explicitly that $S_{\cbm}(\Phi)$ is
 achieved with a maximally entangled state and that it decreases 
 monotonically with $\mu$.  Numerical work \cite{DSS}
 shows that $S_{\cbm}(\Phi)$ changes from positive to negative at $\mu= 0.74592$,
 which is also the cut-off for $C_Q(\Phi) = 0 $.

The Werner-Holevo  channel \cite{WH} is
$\Phi_{\rm WH}(\rho) = \tfrac{1}{d-1} \big[  (\tr \rho) \rmi - \rho^T \big]$.
One finds that   $\gamma_{12}$ has exactly $\binom{d}{2}$
non-zero eigenvalues $\tfrac{1}{ d-1} (a_j^2 + a_k^2)$ with $j < k$ and
$a_j^2$ the eigenvalues of $\gamma_1$.  One can then use the
 concavity of $- x \log x$ to show that 
 $ S(\gamma_{12}) \geq  S(\gamma_1) + \log \tfrac{d-1}{2}$,
which implies that $S_\cbm(\Phi_{\rm WH}) = \log \tfrac{d-1}{2}$
   is achieved with a maximally entangled input.  
Moreover, $S_\cbm(\Phi_{\rm WH}) =  -1$ for $d = 2$,
 and $S_\cbm(\Phi_{\rm WH}) = 0$ for $d = 3$.   
 One  can also use the covariance property
   $\Phi_{\rm WH}(U \rho U^*) = \ovb{U} \Phi_{\rm WH}(\rho) U^T$
   and Lemma~2 of \cite{J} to see that  $\omega_p(\Phi_{\rm WH})$
   is achieved with a maximally entangled state, and verify that
 \be
   \omega_p(\Phi_{\rm WH}) = \big( \tfrac{2}{d-1}  \big)^{1- \frac{1}{p}} >
       \big( \tfrac{1}{d-1}  \big)^{1- \frac{1}{p}} = \nu_p(\Phi_{\rm WH}).
 \ee
This gives another example for which the CB norm is strictly greater than
$\norm{ \Phi}_{1 \raw p}$.

However, the CB norm is not always attained on a maximally entangled state.
Consider for example the non-unital qubit map 
$\Phi(\rho) = \lambda \rho + \big({(1 - \lambda) \over 2} I + {t \over 2} \sigma_3 \big)\, \tr \rho $,
and the one-parameter family of pure bipartite states
$| \psi \ket_a = \sqrt{a} \, | 00 \ket + \sqrt{1-a}\, | 11 \ket $ where $0 \leq a \leq 1$.
In this case
\be
\gamma_{12} & = & (I \ot \Phi) (| \psi \ket_a \, \bra \psi |) \nonumber \\
& = &
\frac{1}{2} \pmx
a(1 + t + \lambda) & 0 & 0 & 2 \lambda \sqrt{a(1-a)} \nonumber \\
0 & (1-a) (1+t-\lambda) &0 & 0 \\
0 & 0& a(1-t-\lambda) &0 \\
2 \lambda \sqrt{a(1-a)} & 0 & 0 & (1-a) (1 -t + \lambda)
\emx
\ee
Numerical computations show that for $p > 1$,
$\frac{ \norm{ \gamma_{12}}_p}{ \norm{ \gamma_1}_p}$ 
is maximized at values $a > 1/2$ when $t > 0$, and values $a < 1/2$ when
$t < 0$. Since the state $| \psi \ket_{a}$ is maximally entangled only when $a = 1/2$,
this demonstrates that the CB norm $\omega_p(\Phi)$ is achieved
at a non-maximally entangled state for this family of maps.

\subsection{Entanglement breaking and preservation}   \label{sect:EBT}

The class of channels for which $(\id \ot \Phi) (\rho)$ is separable for any input
is called entanglement breaking (EB).   Those which are also trace preserving are
denoted EBT.   These maps were introduced in \cite{Hv2}  by Holevo who wrote
them in the form $\Phi(\rho) = \sum_k R_k \tr \rho E_k$ with each $R_k$ a
density matrix and $\{ E_k \}$ a POVM, i.e., $E_k \geq 0$ and $\sum_k E_k = I$.
They were studied in \cite{HSR} where several equivalent conditions were proved.
The next result shows that EBT channels always have positive minimal CB entropy.
Therefore, a channel for which $S_{\cbm}(\Phi) $ is negative always preserves
some entanglement.
\begin{lemma} \label{lemm:EBT}
If $\Phi: M_m \mapsto M_n$ is an EBT map, then for all $p \geq 1$ and 
positive semi-definite $Q \in M_n \ot M_m$, 
\be
\norm{( \id_n \ot \Phi)(Q)}_p \leq  \norm{ \trp_2 \, Q}_p = \norm{Q_1}_p
\ee
\end{lemma}
\begin{thm} \label{thm:EBT}
If $\Phi$ is an EBT map, then  $\omega_p(\Phi) \leq 1$ and
$S_{\cbm}(\Phi) $ is positive.
\end{thm}
Theorem~\ref{thm:EBT} follows immediately from Lemma~\ref{lemm:EBT} 
and Theorem~\ref{thm:CB2} of Section~\ref{sect:CBpf}.
The converse does not holds, i.e.,  $S_{\cbm}(\Phi) \geq 0$ does not imply
that $\Phi$ is EBT.   For the
depolarizing channel,  it is known \cite{RuskEBT} that $\Omega_{\alpha}$
 is EBT if and only if $|\alpha| \leq \tfrac{1}{3}$; however, as reported above,
$S_{\cbm}(\Omega_{\alpha}) > 0$  for $0 < \alpha < 0.74592$.   
For $d > 3$, the WH channel also
has positive CB entropy, although it can not break all entanglement
 because it is known \cite{WH}
 that $\nu_p(\Phi_{\rm WH})$ is not multiplicative for sufficiently large $p$.

 The  proof of Lemma~\ref{lemm:EBT}   is similar to King's argument \cite{King4} 
for showing  
multiplicativity of the maximal $p$-norm for EBT maps, 
and is based on the following inequality due to Lieb  and Thirring \cite{LT}  
  \be  \label{LT}    
     \tr (C^{\dag} D C) \leq \tr (CC^{\dag})^p \, D^p
 \ee
for $p \geq 1$ and $D > 0$ positive semi-definite.\footnote{The proof 
in the Appendix of \cite{LT} is based on the concavity of 
$A \mapsto \tr (B A^{1/m} B)^m$ for $m \geq 1$ and $A,B \geq 0$.
This was first proved by Epstein~\cite{Ep}; it is also a special case of
Lemma 1.14 in \cite{PisLp}, which is proved using complex interpolation
in the operator space framework.
   Araki  \cite{Ak} gave another  proof of   \eqref{LT}, and 
a simple proof based on H\"older's inequality  
was given by Simon in Theorem I.4.9 of \cite{Simon}.   }

\pf{Lemma~\ref{lemm:EBT}}  
By assumption, we can write
$\Phi(\rho) = \sum_k R_k \tr \rho E_k$ with each $R_k$ a
density matrix and $\{ E_k \}$ a POVM.   Then
\be
    ( \id_n \ot \Phi)(Q) & = & \sum_{k = 1}^\kappa [\trp_2 \, (\rmi \ot X_k)Q] \ot  R_k  \nn \\
      & = & \sum_{k = 1}^\kappa G_k \ot  R_k 
\ee
where $G_k = \sum_k [\trp_2 \, (\rmi \ot X_k)Q] $.
Note that
\be
   \trp_2 \, Q =  \sum_{k = 1}^\kappa [\trp_2 \, (\rmi \ot X_k)Q]  =  \sum_{k = 1}^\kappa G_k
   \ee
  With $|e_k\ket$ 
    the canonical basis in ${\bf C}_\kappa$ we define the following matrices
    in $M_\kappa \ot M_n \ot M_n$.
  \be
  R \, = \,  \sum_k \proj{e_k} \ot \rmi_n  \ot R_k \, = \,   \pmx \rmi_n \ot  R_1   & 0 & \ldots & 0 \\
             0 &  \rmi_n \ot R_2 & \ldots & 0 \\
             \vdots & ~ & \ddots &   \vdots \\
             0 & \ldots & 0 & \rmi_n \ot R_\kappa \emx
  \ee 
  and
  \be
  V    \, = \,   \wtd{V} \ot \rmi_n \, = \,  \sum_k | e_k \kb e_1 | \ot  G_k^{1/2} \ot \rmi_n  =
   \pmx  \sqrt{G_1}   & 0 & \ldots & 0 \\
              \sqrt{G_2 }     & 0 & \ldots & 0 \\
             \vdots & \vdots & ~ &   \vdots \\
             \sqrt{G_{\kappa} }   & 0 & \ldots & 0 \emx \ot \rmi_n
                  \ee
     where we adopt the convention of   using the subscripts $3,1,2$ for 
        $M_\kappa, M_n, M_n$  
     respectively so that the partial traces $\trp_1$ and $\trp_2$ retain
    their original meaning.  
It follows that
    \be
     \proj{e_1} \ot ( \id_n \ot \Phi)(Q) = V^{\dag} R V.
    \ee
    Applying \eqref{LT} one finds
    \be
       \norm{( \id_n \ot \Phi)(Q)}_p^p & = &  \tr (V^{\dag} R V)^p ~ = ~  \trp_{312} \, (V^{\dag} R V)^p \nn  \\
          &  \leq &  \trp_{312} \, (VV^{\dag} )^p \, R^p \\ \nn
          & = & \sum_k \trp_{12} \, [(VV^{\dag} )^p]_{kk} (\rmi_n \ot R_k)^p \\
          & = & \sum_k \trp_1 \, [(\wtd{V}\wtd{V}^{\dag} )^p]_{kk}  \trp_2 \, (R_k)^p 
    \ee
    where $[(\wtd{V}\wtd{V}^{\dag} )^p]_{kk} = 
      \trp_3 \, (\wtd{V}\wtd{V}^{\dag} )^p (\proj{e_k} \ot \rmi_n)$ is the
    $k$-th block on the diagonal of $(\wtd{V}\wtd{V}^{\dag} )^p$ and
    $ [(VV^{\dag} )^p]_{kk} = [(\wtd{V}\wtd{V}^{\dag} )^p]_{kk} \ot \rmi_n$.   
      Since $R_k$ is a density matrix,
    $ \trp_2 \, (R_k)^p \leq 1$.  (In fact, we could assume wlog that
    $R_k = \proj{ \theta_k}$ so that $R_k^p = R_k$ and $ \trp_2 \, (R_k)^p = 1$.)
    Therefore,
    \be
     \norm{( \id_n \ot \Phi)(Q)}_p^p & \leq &  \sum_k \trp_1 \, [(\wtd{V} \wtd{V}^{\dag} )^p]_{kk} \nn  \\
      & = & \trp_{31} \, (\wtd{V} \wtd{V}^{\dag})^p  ~ = ~  \trp_{31} (\wtd{V}^{\dag} \wtd{V})^p \\  \nn 
      & = & \sum_k  \trp_1 \, G_k  ~ = ~ \trp_2 \, Q   \qquad \qed
    \ee
    
\subsection{Operational interpretation}

Recently Horodecki,  Oppenheim and  Winter \cite{HOW} (HOW) obtained
results which give an important operational meaning to quantum conditional
information, consistent with both positive or negative values.    Applying their
results to the expression $S_{\cbm}(\Phi) = S(\gamma_{AB}) - S(\gamma_A)$ 
with $\gamma_{AB} =  (\id \ot \Phi) \big(|\psi \kb \psi| \big) $ where
$|\psi\ket$ is the minimizer in \eqref{cbmdef} gives the following interpretation:
\begin{itemize}

\item A channel for which $S_{\cbm}(\Phi) > 0$ always breaks enough entanglement
so that some EPR pairs must be added to enable Alice to  transfer   her
information to Bob. 

\item A channel for which $S_{\cbm}(\Phi) < 0$  leaves enough entanglement
in the optimal state so that some EPR pairs remain after Alice has transferred her
information to Bob. 

\end{itemize}
For example, as discussed in Section~\ref{sect:exam} the depolarizing channel is 
entanglement breaking for
$\mu \in [- \tfrac{1}{3},\tfrac{1}{3}]$; for $\mu \in  (\tfrac{1}{3},0.74592)$ it always
breaks enough entanglement to require input of EPR pairs to transfer Bob's
corrupted state back to Alice; and for $\mu > 0.74592$ maximally entangled states
retain enough entanglement to allow the distillation of EPR pairs after Bob's
corrupted information is transferred to Alice.

Note, however, that the  HOW interpretation \cite{HOW} is an asymptotic result in the sense
that it is are based on the assumption of the availability of the tensor product
state $\gamma_{AB}^{\ot n}$ with $n$ arbitrarily large, and is
related to the  ``entanglement of assistance'' \cite{SVW} which is known
not to be additive.    
One would also like to have an interpretation of the additivity of $S_{\cbm}(\Phi) $
so that the ``one-shot'' formula $- S_{\cbm}(\Phi) $ represents the capacity
of an asymptotic process which  is not enhanced by entangled inputs.     
Thus far,  the only scenarios for which we have found this to be true
seem extremely contrived and artificial.    

\section{Entropy Inequalities} \label{sect:ent}

In this section, we show that operator space methods can be used to
give a new proof of SSA \eqref{SSA}.    Although the
strategy is  straightforward, it requires some rather lengthy and tedious
bounds on derivatives and norms.    Our purpose is not to give another
proof of SSA, but to demonstrate the fundamental role of Minkowski-type
inequalities and provide some information on the behavior  of the
$\norm{~}_{(1,p)}$ near $p = 1$.

Differentiation of inequalities of the type found in Section~\ref{sect:fubmink}
often yields entropy inequalities.   The procedure is as follows.
Consider an inequality of the form $g_L(p) \leq g_R(p)$ valid for $p \geq 1$
  which becomes an equality at $p = 1$.   Then the function
  $g(p) = g_R(p) - g_L(p) \geq 0$ for $p \geq 1$ and $g(1) = 0$.  This implies
  that the right derivative  $g^{\prime}(1+) \geq 0$ or, equivalently, that
  $g_L^{\prime}(1+) \leq g_R^{\prime}(1+)$.  

  Applying this to \eqref{mink.qp} yields
  \be  \label{WSA}
      - S(Q_1) \leq  -S(Q_{12}) + S(Q_2)
  \ee
which is the well-known subadditivity inequality $ S(Q_{12}) \leq  S(Q_1) + S(Q_2)$.
Applying the same principle to conjecture \eqref{mink3} yields
 \be \label{SSA2}
      - S(Q_{23}) + S(Q_3) \leq  -S(Q_{123}) + S(Q_{13})
  \ee
which is equivalent to strong subadditivity  \eqref{SSA}.
    (Carlen and Lieb \cite{CL} observed
  that the reverse of  \eqref{mink3} holds when $t \leq 1$ and used the
  corresponding left
  derivative inequality $g_L^{\prime}(1-) \geq g_R^{\prime}(1-)$
  to obtain another proof of SSA.)

These entropy inequalities can also be obtained by differentiating the corresponding
CB Minkowski inequalities  \eqref{mink}  and  \eqref{cb3mink}.    We will need
the following.
\begin{thm}   \label{thm:dff2}
For any $X = X_{12}$ in $M_m \ot M_n$, with $X \geq 0$ and $\tr X = 1$.
\be\label{eqn:dff2}
  \tfrac{d~}{dp} \, \norm{X_{12}}_{(1,p)}^p \, \big|_{p = 1} & =   -S(X_{12}) + S(X_1).
\ee
\end{thm}

Before proving this result, observe that
\eqref{fm4b} implies $\norm{W_{12}}_{(1,p)} = \norm{W_2}_p$
and   \eqref{trip.1p1.13} implies
    $\norm{W_{132}}_{(1,p,1)} = \norm{W_{13}}_{(1,p)}  $.     Then, when $q = 1$,
    the inequalities \eqref{mink}  and  \eqref{cb3mink} imply
    \be   \label{minkq=1}
     \norm{W_2}_p^p  & \leq & \norm{W_{12}}_{(1,p)}^p \\  \label{mink3q=1}
     \norm{W_{13}}_{(1,p)}^p & \leq & \norm{W_{123}}_{(1,1,p)}^p.
    \ee
Now, under the assumption that $W_{123} > 0$ and $\tr W_{123} = 1$,
Theorem~\ref{thm:dff2} implies
\bee
\tfrac{d~}{dp} \, \norm{W_{21}}_{(1,p)}^p\, \big|_{p = 1} & =  &  -S(W_2)     \\
  \tfrac{d~}{dp} \, \norm{W_{12}}_{(1,p)}^p \, \big|_{p = 1} & = &   -S(W_{12}) + S(W_1) \\
    \tfrac{d~}{dp} \, \norm{W_{123}}_{(1,1,p)}^p \, \big|_{p = 1}  & = &   -S(W_{123}) + S(W_{12}) \\
      \tfrac{d~}{dp} \, \norm{W_{13}}_{(1,p)}^p \, \big|_{p = 1}  & =  & -S(W_{1 3}) + S(W_{1})
\eee
Then usual subadditivity and SSA inequalities,  \eqref{WSA} and \eqref{SSA2}
then follow from  the principle, $g_L^{\prime}(1+) \leq g_R^{\prime}(1+)$, above
and \eqref{minkq=1} and  \eqref{mink3q=1} respectively.

\pf {Theorem~\ref{thm:dff2}}  The basic strategy is similar to that in Section~\ref{sect:CBpf}, but
requires some additional details.
Let  $X_1 = \trp_2 \, X$ and let $Q$ denote the orthogonal projection onto  $\ker(X_1)$. 
Since $Q X_1 Q = 0$, it follows that
$\tr (Q \ot  \rmi_n) X (Q \ot \rmi_n) = 0$.   
Since  $X$ is positive semi-definite this implies that 
$X = ((\rmi_m -Q) \ot  \rmi_n) X ((\rmi_m -Q) \ot \rmi_n)$.
For fixed $X$ the functions
  \be  \label{def:v}
   v(p,B) 
  & =   & X^{1/2} (B^{\frac{1}{p} -1} \ot \rmi_n) \, X^{1/2},  \quad
\text{and}  \\
w(p,B) & = & X_{1}^{\haf} \, B^{\frac{1}{p} -1} \, X_{1}^{\haf}.  \label{def:w}
\ee
are well-defined for $p > 1$, and  $B \in \beta(X_1)$  where
 $\beta(X_1) =  \{ B \in {\cal D} : \ker(B) \subset \ker(X_1) \}$.
 Since  $\norm{ (B^{- \haf(1 - \frac{1}{p})} \ot \rmi_n) \, X \,
      (B^{- \haf(1 - \frac{1}{p})} \ot \rmi_n) }_p  
      =   \norm{ X^{\haf} (B^{-\haf} B^{\frac{1}{p}} B^{-\haf} \ot \rmi_n) X^{\haf}}_p$,
it follows from \eqref{fm5c} and the remarks above that
 \be
 \norm{X_{12}}_{(1,p)} =   \inf_{B \in \cal D}  \norm{ v(p,B) }_p =
      \inf_{B \in \beta(X_1)} \norm{ v(p,B) }_p .
 \ee
The set of density matrices ${\cal D}$ is compact, and $\norm{ v(p,B) }_p$ is bounded below
and continuous, hence for each $p > 1$ there is a (i.e., at least one) density matrix $B(p)$
which minimizes $\norm{ v(p,B) }_p$, so that
\be
\norm{X_{12}}_{(1,p)} = \norm{ v(p,B(p)) }_p
\ee
Since $p  > 1$ and  $B(p)$ is a density matrix,   $B(p)^{- 1 + \frac{1}{p}} > \rmi_m$ which implies
\be  \label{Xbd}
    v(p,B) \geq X \qquad \text{and} \qquad  w(p,B) \geq X_1.
\ee
Furthermore,
\be\label{v.bound1}
1 = \tr X \leq \tr v(p,B(p)) \leq (m n)^{\frac{1}{p} - 1} \norm{ v(p,B(p)) }_p
\ee
(where the last inequality uses $ \norm{A}_1 \leq d^{1 - \frac{1}{p}} \norm{A}_p$
for any positive semi-definite $d \times d$ matrix $A$ and any $p \geq 1$).
Replacing $B(p)$ by another density matrix cannot decrease $\norm{ v(p,B(p)) }_p$, hence
\be\label{v.bound2}
\norm{ v(p,B(p)) }_p \leq  \norm{ v \big(p,  \frac{1}{m} \rmi_m \big) }_p = m^{1 - \frac{1}{p}} \, \norm{ X }_p \leq 
m^{1 - \frac{1}{p}}
\ee
Combining (\ref{v.bound1}) and (\ref{v.bound2}) shows that
\be\label{lim.v}
\lim_{p \raw 1+} \, \tr (v(p,B(p)) - X) = 0,
\ee
and, together with  \eqref{Xbd} implies   that $v(p,B(p))  \rightarrow X$.  
Also, for any $B \in \beta(X_1)$,
 \be  \label{tr.equal}
\tr v(p,B) & = & \trp_{12}  \, X_{12} \,  [B]^{\frac{1}{p} - 1} \ot \rmi_n)   \nn \\
& = & \tr  X_1 \, [B]^{\frac{1}{p} - 1}  =  \tr w(p,B)
\ee
so that $\lim_{p \raw 1+} \, \tr (w(p,B(p)) - X_1) = 0$ and $w(p,B(p))  \rightarrow X_1$.

Writing out the derivative on the left side of \eqref{eqn:dff2}, we see that
 we need to show that
\be\label{goal1}
\lim_{p \rightarrow 1+} \, \frac{1}{p-1} \, \Big( \tr v(p,B(p))^p -1 \Big) = - S(X) + S(X_1)
\ee
First note that for $p > 1$.
\be\label{up.bound}
\frac{1}{p-1} \, \Big( \tr v(p,B(p))^p -1 \Big)
\leq \frac{1}{p-1} \, \Big( \tr v(p, X_1)^p - 1 \Big),
\ee
and a direct calculation shows that the right side of (\ref{up.bound}) converges to 
 $- S(X) + S(X_1)$  as $p \raw 1+$. Hence to prove (\ref{goal1})
it is sufficient to show that
\be\label{goal2}
\liminf _{p \raw 1+} \frac{1}{p-1} \, \Big( \tr v(p,B(p))^p -1 \Big)
\geq - S(X) + S(X_1)
\ee
 H\"older's inequality implies
\be
1 = \norm{X_1}_1 & = & \norm{B(p)^{\haf - \frac{1}{2p}} \,
\Big(B(p)^{\frac{1}{2p} -\haf} \, X_1 \, B(p)^{\frac{1}{2p} -\haf}\Big) \, B(p)^{\haf - \frac{1}{2p}}}_{1} \nn \\
& \leq & \norm{B(p)^{\haf - \frac{1}{2p}}}_{2p/p-1}^{2} \,\,\, \norm{B(p)^{\frac{1}{2p} -\haf} \, X_1 \, B(p)^{\frac{1}{2p} -\haf}}_{p} \nn \\
& = & \norm{w(p,B(p))}_{p}
\ee
Combining this with \eqref{tr.equal} gives a bound on the numerator on the left in  \eqref{goal1}
\be\label{lower1}
  \tr v(p,B(p))^p  - 1      & \geq &   \tr v(p,B(p))^p - \tr v(p,B(p))   \nn \\
  & ~ & ~ - \Big[\tr w(p,B(p))^p - \tr w(p,B(p)) \Big]   \ee
 
The mean value theorem for the function $g(p) = x^p$ implies that
for some $p_1, p_2 \in [1,p]$
\bsq \label{mean}   \be 
\frac{1}{p-1} \, \Big(\tr v(p,B(p))^p - \tr v(p,B(p))\Big) & = & \tr v(p,B(p))^{p_1} \, \log v(p,B(p))  \\
\frac{1}{p-1} \, \Big(\tr w(p,B(p))^p - \tr w(p,B(p))\Big) & = & \tr w(p,B(p))^{p_2} \, \log w(p,B(p)) \label{meanb}
\ee \esq
The convergence in \eqref{lim.v} and following \eqref{tr.equal} imply
\bsq   \label{lim} \be\label{lim1}
\lim_{p \rightarrow 1} \, \tr v(p,B(p))^{p_1} \, \log v(p,B(p)) & = & - S(X) \\
 \label{lim2}
\lim_{p \rightarrow 1} \, \tr w(p,B(p))^{p_2} \, \log w(p,B(p)) & = &  - S(X_1).
\ee  \esq
Combining \eqref{mean}, \eqref{lim}and  (\ref{lower1}) gives (\ref{goal2}). \quad  \qed

 \rmk The proof above relies on the convergence of
 $ \lim_{p \raw 1+} X^{\half} (B^{\frac{1}{p} -1} \ot \rmi_n) \, X^{\half} = X$ and 
$ \lim_{p \raw 1+} X^{\half}X_1^{\half} B^{\frac{1}{p} -1}  \, X_1^{\half} = X_1$,
but tells us nothing at all about the behavior of $B(p)$  as $p \raw 1+$.
 By making a few changes at the end of this proof and exploiting Klein's
 inequality, we can also show
 that $\lim_{p \raw 1+} B(p) = X_1$.  
 
 Klein's inequality \cite{Klein,NC} states that 
\be \label{Klein}
\tr A \log A -  \tr A \log B  \geq \tr (A - B) 
\ee
 with equality in the case $\tr A = \tr B$ if and only if $A = B$.
 
 Now, replace \eqref{lower1} by 
\be  \label{lower1K}
\tr v(p,B(p))^p  - 1 & = & 
\tr v(p,B(p))^p - \tr v(p,B(p)) + (  \tr w(p,B(p)) - 1) .
\ee
Then use the mean value theorem for the function 
$g_2(p) = y^{\frac{1}{p}}$ to replace \eqref{meanb} by
\be  \label{lower2K}
\frac{1}{p-1} \, \Big(\tr w(p,B(p))  - 1 \Big) & = & - \frac{1}{\wtd{p}^2} 
     \tr X_1^{\haf} \, B(p)^{\frac {1}{\wtd{p}} -1)} \log  B(p)  \,    X_1^{\haf}   .
\ee
We could use \eqref{Klein}
with 
$A = B(p)^{- \haf(1 - \frac{1}{\wtd{p}})}  X_1 B(p)^{- \haf(1 - \frac{1}{\wtd{p}})}  $ 
together with the fact that $A$ and $w(\wtd{p},B(p))$ have the same non-zero
 eigenvalues  to bound the right side of \eqref{lower2K} below
by  $-    \frac{1}{\wtd{p}^2} S[w(\wtd{p},B(p)] +  \tr  w(\wtd{p},B(p)) - 1   $. 
However, because 
 $1 < \wtd{p} < p$ implies $B^{1/\wtd{p}} > B^{1/p } $, we cannot extend
   \eqref{v.bound2} and \eqref{tr.equal} to conclude that this converges to $S(X_1)$.

Instead, we first observe that the compactness of the set of density matrices
${\cal D}$ implies that we can find a sequence $p_k \raw 1+$ such that
$ \norm{X_{12}}_{(1,p)} =     \norm{ v(p_k,B(p_k)) }_{p_k}$ and
$B_k \raw B^* \in {\cal D}$.   If $B^*$ is not in $\beta(X_1)$, then the
right side of the first line of \eqref{lower2K} $\raw + \infty$ giving a contradiction
with \eqref{up.bound}.  Hence $B^* \in \beta(X_1)$.  Therefore,  
\eqref{lower2K} and \eqref{Klein} imply
 \be
  \lim_{p_k \raw \infty}  \frac{1}{p_k-1} \, \Big(\tr w(p_k,B(p_k))  - 1 \Big)
     & = & - \tr X_1 \log B^*  
     \geq S(X_1).
     \ee
Inserting this in \eqref{lower1K} yields
\be  \label{kbnd}
  \lim_{p_k \raw \infty}  \frac{1}{p_k-1} \, \Big(\tr v(p_k,B(p_k))^p  - 1 \Big)
     & = &   -S(X_{12})   - \tr X_1 \log B^*  \nn \\
     & \geq &  -S(X_{12}) + S(X_1)  .
     \ee     
 Combining these results with  \eqref{up.bound}, we conclude that
 equality holds in \eqref{kbnd} and that  
 \be
   - \tr X_1 \log B^* = S(X_1) = - \tr X_1 \log X_1 .
 \ee
We can now use the condition for equality in \eqref{Klein}
   to conclude that $B^* = X_1$.    Since this is
true for the limit of {\em any} convergent sequence of minimizers
$B(p_k)$ with $p_k \raw 1$, we have also proved the following 
which is of independent
interest.
\begin{cor}
For   $X  \in M_m \ot M_n$ with $X \geq 0$ and $\tr X = 1$ and $p \in (1,2]$,
let $B(p) \in {\cal D}$ minimize $\norm{X}_{(1,p)}  $, i.e.,
    $\norm{X^{\frac{1}{2}} (B^{\frac{1}{p} -1} \ot \rmi_n) \, X^{\frac{1}{2}}}_p = \norm{X}_{(1,p)}  $.
  Then $\ds{\lim_{p \raw 1+} B(p) = X_1} \equiv \trp_2 \, X$.
\end{cor}

  \medskip


   \noindent{\bf Acknowledgments:}
The work of M.J. was supported in part by National Science Foundation grant DMS-0301116.
  The work of C.K.  was  supported in part by the
National Science Foundation under grant DMS-0400426.
The work of M.B.R. was supported in part    by  the National Security 
Agency (NSA) and Advanced Research and Development Activity (ARDA) 
under Army Research Office (ARO) contract number  DAAD19-02-1-0065; 
and  by the National Science Foundation under Grant 
DMS-0314228.    

This work had its genesis in a workshop  in 2002 at the Pacific
Institute for the Mathematical Sciences at which M.J. and M.B.R. 
participated.   Part of this work was done while I.D. and M.B.R. were
visiting the Isaac Newton Institute.    The authors are grateful to these
institutions for their hospitality and support.   
Finally, we thank Professor Andreas Winter for discussions about
possible interpretations of  $S_{\cbm}(\Phi)$.



\appendix  

\section{Purification}   \label{app}

To make this paper self-contained and accessible to people in
fields other than quantum information we summarize the results 
needed to prove Lemma~\ref{lemm:igor}.   

Any density matrix in ${\cal D}_d$ can be written in terms of its
spectral decomposition (restricted to $[\ker(\gamma)]^\perp$) as
$   \gamma = \sum_{k = 1}^m \lambda_k  \, \proj{\phi_k }$
where each  eigenvalue $\lambda_k > 0$   and counted in terms of
its multiplicity so that the eigenvectors $\{ |\phi_k \ket \}$ are orthonormal.
If we then let $\{ |\chi_k \ket \}$ be any orthonormal basis of ${\bf C}^m$
and define $  |\Psi\ket  \in  {\bf C}^d \ot  {\bf C}^m$ as
\be  \label{pur}
  |\Psi\ket =  \sum_{k = 1}^m   \sqrt{ \lambda_k} \, | \phi_k \ot |\chi_k \ket .
  \ee
then $    \gamma = \trp_2 \, \proj{ \Psi}$ and \eqref{pur}
is called a {\em purification} of $\gamma$. 

Conversely, given a normalized vector $  |\Psi\ket  \in  {\bf C}^n \ot  {\bf C}^m$,
it is a straightforward consequence of the singular value decomposition
that  $  |\Psi\ket $ can be written in the form
\be  \label{svd}
  |\Psi\ket =  \sum_k  \mu_k  \, | \phi_k \ot |\chi_k \ket
  \ee
  with $\{ | \phi_k\}$ and $\{ |\chi_k \ket \}$ orthonormal sets in 
 $ {\bf C}^n$ and ${\bf C}^m$ respectively.    (This is often called the ``Schmidt
 decomposition'' in  quantum information theory.
 For details and some history
 see Appendix~A of \cite{KR1}.)  It follows from
 \eqref{svd}  that the reduced
 density matrices $\gamma_1 =  \trp_2   \proj{ \Psi} $ and
 $\gamma_2 =  \trp_1   \proj{ \Psi} $ have the same non-zero eigenvalues.
 Although our interest here is for  $\hil =  {\bf C}^m$, these results
 extend to infinite dimensions and yield the following   
 \begin{cor} 
 When $ |\Psi_{AB} \ket $ is a bipartite pure state in $\hil_A \ot \hil_B$,
 then its reduced density matrices $\gamma_A =  \trp_B \proj{ \Psi} $ and
 $\gamma_B =  \trp_A  \proj{ \Psi} $ have the same entropy, i.e.,
$\nolinebreak{    S(\gamma_A) = S(\gamma_B)}$.
 \end{cor}
 

{~~}

\end{document}